\documentclass{aa}  

\usepackage{upgreek} 
\usepackage{multirow}
\usepackage{graphicx}
\usepackage{enumitem}
\usepackage[dvipsnames]{xcolor}
\usepackage{subcaption}
\usepackage{hyperref}
\hypersetup{colorlinks=True, allcolors=blue}
\def\le{\leqslant}
\def\lms{\rm \log (M_*/M_\odot)}
\def\Reff{\rm R_{\mathrm{EFF}}}
\begin{document}

   \title{Galaxy Size and Mass Build-up in the First 2 Gyrs of Cosmic History from Multi-Wavelength JWST NIRCam Imaging}

   \titlerunning{The size--mass relation at $z>3$}
   \authorrunning{N. Allen et al.}

   \author{Natalie~Allen\inst{1,2}, 
          Pascal~A.~Oesch\inst{1,2,3},
          Sune~Toft\inst{1,2},
          Jasleen~Matharu\inst{1,2},
          Conor~J.~R.~McPartland\inst{1,2},
          Andrea~Weibel\inst{3},
          Gabe~Brammer\inst{1,2},
          Rebecca~A.~A.~Bowler\inst{4},
          Kei~Ito\inst{1,5,6},  
          Rashmi~Gottumukkala\inst{1,2},
          Francesca~Rizzo\inst{7},
          Francesco~Valentino\inst{8,1}, 
          Rohan~G.~Varadaraj\inst{9},
          John~R.~Weaver\inst{10}, 
          Katherine~E.~Whitaker\inst{11,1}
          }

   \institute{Cosmic Dawn Centre (DAWN), Copenhagen, Denmark \\ \email{natalie.allen@nbi.ku.dk}
        \and
        Niels Bohr Institute, University of Copenhagen, Jagtvej 128, DK-2200 Copenhagen N, Denmark 
        \and
        Department of Astronomy, University of Geneva, Chemin Pegasi 51, 1290 Versoix, Switzerland
        \and 
        Jodrell Bank Centre for Astrophysics, University of Manchester, Oxford Road, Manchester, M13 9PL, UK
        \and 
        DTU-Space, Elektrovej, Building 328 , 2800, Kgs. Lyngby, Denmark
        \and
        University of Tokyo, Hongo, Bunkyo-Ku, Tokyo, 113-0033, Japan 
        \and 
        Kapteyn Astronomical Institute, University of Groningen, Landleven 12, 9747 AD, Groningen, The Netherlands
        \and
        European Southern Observatory, Karl-Schwarzschild-Str. 2, D-85748 Garching bei Munchen, Germany 
        \and
        Sub-department of Astrophysics, University of Oxford, Denys Wilkinson Building, Keble Road, Oxford, OX1 3RH, UK
        \and 
        Department of Astronomy, University of Massachusetts, Amherst, MA 01003, USA }

   \date{Received ...; accepted ...}

\abstract{The evolution of galaxy sizes in different wavelengths provides unique insights on galaxy build-up across cosmic epochs. Such measurements can now finally be done at $z>3$ thanks to the exquisite spatial resolution and multi-wavelength capability of the JWST. With the public data from the CEERS, PRIMER-UDS, and PRIMER-COSMOS surveys, we measure the sizes of $\sim 3500$ star-forming galaxies at $3\le z<9$, in 7 NIRCam bands using the multi-wavelength model fitting code \texttt{GalfitM}. The size--mass relation is measured in four redshift bins, across all NIRCam bands. We find that, the slope and intrinsic scatter of the rest-optical size--mass relation are constant across this redshift range and consistent with previous HST-based studies at low-z.  When comparing the relations across different wavelengths, the average rest-optical and rest-UV relations are consistent with each other up to $z=6$, but the intrinsic scatter is largest in rest-UV wavelengths compared to rest-optical and redder bands. This behaviour is independent of redshift and we speculate that it is driven by bursty star-formation in $z>4$ galaxies. Additionally, for $3\le z<4$ star-forming galaxies at $\rm M_* > 10^{10} M_{\odot}$, we find smaller rest-$\rm 1\rm\,\mu m$ sizes in comparison to rest-optical (and rest-UV) sizes, suggestive of colour gradients.
When comparing to simulations, we find agreement over $\rm M_* \approx  10^{9} - 10^{10} M_{\odot}$ but beyond this mass, the observed size--mass relation is significantly steeper. Our results show the power of JWST/NIRCam to provide new constraints on galaxy formation models.
}

\keywords{Galaxies: structure -- Galaxies: high-redshift -- Galaxies: evolution -- Galaxies: statistics -- Galaxies: morphology}

   \maketitle

\section{Introduction}

\setlist{nolistsep}

The evolution of galaxy sizes with redshift provides unique constraints on the build-up of galaxies across cosmic history. Galaxies form as gas accretes and cools within Dark Matter (DM) halos, leading to a correlation between galaxy sizes and their host structures, which scale as $\rm R_{ virial} \propto M_{\rm halo}^{1/3}$ (\citealt{Huang2017Relations0z3, Mo1998TheDiscs, Fall1980FormationHaloes}). Where $\rm R_{\rm virial}$ is the viral radius and $\rm M_{halo}$ is the mass of the DM halo. It is expected that the sizes of galaxies are related to their DM halos and therefore galaxy sizes will follow a similar power law relation between their effective radius, $\Reff$ and their stellar mass, $\rm M_*$.
\begin{equation}
    \rm \Reff \propto M^{\alpha}_{*}
\end{equation}

Throughout their lifetime galaxies also experience a variety of physical processes such as star formation, stellar and active galactic nuclei (AGN) feedback, mergers, environment and gas accretion which all impact and change their structure. Thus, the sizes of galaxies are used as fundamental measurements to probe and understand galaxy evolution, providing insights into how the largest galaxies we see in our local Universe formed. 

The Hubble Space Telescope's (HST) high spatial resolution has enabled the study of rest-UV and rest-optical galaxy sizes up to $z\sim 10$ and $z\sim 3$, respectively, of both star-forming and quiescent galaxy populations. Studies in the rest-frame optical wavelengths find that the stellar size--mass relation of star-forming galaxies is well-described by a single power law given by Equation \ref{equ:log-size-mass} (e.g. \citealt{VanDerWel20143D-HST+CANDELS:3, Mowla2019COSMOS-DASH:CANDELS/3D-HST, Nedkova2021ExtendingCANDELS}).

\begin{equation}
     \rm \log \Reff = \alpha \log\left(\frac{M_*}{M_0}\right) + \log A
    \label{equ:log-size-mass}
\end{equation}

Here, $\rm M_0$ is a normalisation parameter in units of solar mass, $\alpha$ is the slope and $\rm \log A$ is the intercept of the log size--mass relation, i.e., the size of a source at stellar mass $\rm M_0$.
We can also measure the width of the distribution of galaxy sizes, called the intrinsic dispersion of a lognormal ($\sigma_{\log \Reff}$), by including it in our modelling of Equation \ref{equ:log-size-mass}.  

Previous HST-based studies have examined the evolution of the size--mass relation parameters to infer the physical processes influencing galaxy structure. Most of these studies find that the slope and intrinsic scatter of the star-forming rest-optical size--mass relation is constant with redshift (e.g. \citealt{VanDerWel20143D-HST+CANDELS:3, Mowla2019COSMOS-DASH:CANDELS/3D-HST, Nedkova2021ExtendingCANDELS}). For example, \citet{VanDerWel20143D-HST+CANDELS:3} finds that the evolution of the slope and intrinsic scatter between $z=0-3$ are consistent with no evolution with redshift (0.21 and 0.17-19 dex, respectively), while the normalisation changes as $-0.75\log_{10}(1+z) +0.91$. On average, previous works find that star-forming galaxies of mass $\rm 5\times 10^{10} M_\odot$ grow by a factor of 2$\times$ between the redshift range of $0<z<3$. This rate of growth is similar to the dark matter halo growth, supporting the interpretation that the sizes of star-forming galaxies are set by their dark matter halo and the size distribution at a fixed mass is connected to the distribution of dark matter angular momenta (c.f. \citealt{Mo1998TheDiscs, VanDerWel20143D-HST+CANDELS:3}). 

These studies also found that the size evolution depends strongly on galaxy type because assembly histories of star-forming and quiescent galaxies are different. In the standard picture, star-forming galaxies grow through the accretion of cold gas which ignites star-formation, while quiescent galaxies build up their sizes through dry mergers (\citealt{VanDokkum2015FORMINGGALAXIES, Hopkins2008DISSIPATIONSPHEROIDS}). Therefore, it is important to study these two types of galaxy populations separately to disentangle how galaxy sizes are affected by the physical processes driving these different assembly histories.

Pushing such studies to higher redshifts is important to understand the early build-up of galaxies. However, before the launch of the James Webb Space Telescope (JWST), sizes of $z>3$ galaxies could only be  measured in the rest-UV with HST (e.g. \citealt{Bouwens2021SIZESPROGRAM,Oesch2018The,Kawamata2018Size-LuminosityData,Shibuya2015MORPHOLOGIESEVOLUTION,Mosleh2010THEFIELD,Curtis-Lake2016Non-parametric8}). Although, the rest-UV light is sensitive to star formation over the last $\sim100$ Myr which may not be a good representation of the stellar distribution in galaxies at $z>3$. Rest-optical light is more sensitive to the past star-formation history and thus should be a good tracer of the underlying stellar mass.  

Also before the launch of JWST, cosmological and zoom-in simulations predicted measurements of galaxy sizes and their evolution. These models also provide insights into the physical processes driving their evolution and can be used to understand observations of galaxy sizes (e.g. FLARES: \citealt{Lovell2021FirstEvolution, Roper2022FLARESEvolution}, BLUETIDES: \citealt{Feng2015TheReionization,Marshall2021TheReionization},  Illustris-TNG: \citealt{Marinacci2018FirstFields,Nelson2018FirstBimodality,Springel2018FirstClustering,Pillepich2018FirstGalaxies,Naiman2018FirstEuropium,Popping2022The2}, THESAN: \citealt{Kannan2022IntroducingReionization,Garaldi2022TheGalaxies,Smith2022TheReionization,Shen2024TheReionization} and FIRE: \citealt{Hopkins2018FIRE-2Formation, Ma2018SimulatingSizes}.
One surprising result from simulations is that the intrinsic size--luminosity and size--mass relations for galaxies at $z>3$ are described by a power law with a negative slope \citep{Roper2022First5, Costantin2023ExpectationsView, LaChance2024THESIMULATION, Shen2024TheReionization}. In contrast to observations, the simulated galaxies at $\lms > 10$ are \textit{intrinsically} more compact than at lower masses. Only when including the effects of dust does the slope of size--mass (and size--luminosity) relation increase. This dust is found to be centrally compact dust in these massive galaxies, which attenuates emission from the central star-forming core, such that star-formation in the outskirts dominates in simulated observations (\citealt{Roper2022First5, Marshall2021TheReionization, Popping2022The2}). Therefore, the measured sizes in rest-UV and rest-optical may be overestimated in observations.

With the advent of JWST, $z>3$ galaxies can finally be studied in their rest-frame optical light at high spatial resolution, for accurate size measurements across the rest-frame UV to optical spectrum.
In the first two years of science with JWST, there have been numerous studies on the sizes of galaxies at high-z (e.g., \citealt{Ono2023CensusFormation, Ward2023EvolutionCEERS, Morishita2023Enhanced14, Ormerod2023EPOCHSObservations, Varadaraj2024ThePRIMER, Matharu2024ASpectroscopy}). These works find a positive size--mass or size--luminosity relation at $z>3$ in rest-frame UV and rest-optical wavelengths. Some JWST-based rest-UV studies find that star-forming galaxies at $z>4$, in the absolute magnitude range of $-22<M_{UV}<-17$, have rest-optical and rest-UV size ratios of 1 (\citealt{Ono2023CensusFormation,Morishita2023Enhanced14}). This behaviour is in contrast to low-z studies ($z=0-2$) where rest-UV and rest-optical are significantly different, suggesting inside-out growth (\citealt{Ono2023CensusFormation, Matharu2024ASpectroscopy, Shen2024NGDEEPSpectroscopy, Nelson2013THESURVEY, Nelson2016SPATIALLY1.4, Kamieneski2023AreGrowth, VanDerWel20143D-HST+CANDELS:3}). These similar radii at $z>4$ could mean that galaxies are in the initial stages of inside-out growth, dominated by stochastic star formation or dust attenuation is playing a role to make these sizes similar (\citealt{Roper2022FLARESEvolution,Morishita2023Enhanced14}).  

In this paper, we aim to extend the rest-optical size measurements of star-forming galaxies, from HST-based studies, to $z=9$ using high spatial resolution multi-band imaging data from several public JWST/NIRCam surveys. 
Previous observational work also focused their size measurements on one or two bands only. This work differs from these by studying the sizes of galaxies over a range of wavelengths, using several JWST/NIRCam bands. This allows us to understand the stellar mass build up, as well as any biases that come from focusing on measurements in a single band. 

This paper is organised as follows. Section \ref{sec:data_catalogue} describes the catalogue and SED fitting to select the $z>3$ sample. We describe how the sizes are measured and the fitting of the size--mass relation in Section \ref{sec:galfitm}. Section \ref{sec:results} and Section \ref{sec:discussion} present the results and discussion, which are then summarise in Section \ref{sec:conclusion}. 
Throughout this paper, we use a standard flat, cold dark matter cosmology with $\Omega_m = 0.27$, $\Omega_\Lambda = 0.73$ and $H_0 = 70 \mathrm{km s}^{-1} \mathrm{Mpc}^{-1}$. The fluxes and magnitudes used in this paper are specified in the AB system \citep{Oke1983SecondarySpectrophotometry.} and our stellar mass estimates are based on a broken power-law IMF as described in \citet{Eldridge2017BinaryResults} based on \citet{Kroupa1993TheDisc}. If required for comparison, we convert masses based on a \citet{Salpeter1955TheEvolution.} or \citet{Chabrier2003GalacticFunction} IMF to \citealt{Kroupa1993TheDisc} adopting the factors specified in \citet{Madau2014CosmicHistory}.

\section{Data and Catalogue}
\label{sec:data_catalogue}
The imaging data used in this work are the version 7 mosaics of the CEERS, PRIMER-UDS, and PRIMER-COSMOS fields supplied in the DAWN JWST Archive (DJA \footnote{The DAWN JWST Archive (DJA) is a repository of public JWST data, reduced with \texttt{grizli} and \texttt{msaexp}, and released for use by anyone: \url{https://dawn-cph.github.io/dja/index.html}}; \citealt{Valentino2023AnFields}). DJA contains fully reduced mosaics of both the public JWST imaging as well as the ancillary HST data in these fields, using the grism redshift \& line analysis software,  \texttt{grizli} (\citealt{BrammerG.2023Grizli}). In this section, we describe each of these datasets in turn.

\subsection{CEERS}
The Cosmic Evolution Early Release Science Survey (CEERS; Program ID:1345, PI Finkelstein, \citealt{Bagley2022CEERSResults}) is an Early Release Science (ERS) JWST survey. This public survey contains both NIRCam imaging and NIRCam grism $R\sim 1500$ spectroscopy, as well as NIRSpec $R\sim 100$ and $R\sim 1000$ spectroscopy coverage and MIRI Imaging over the Extended Groth Strip (EGS) field. In this paper, we primarily use the NIRCam imaging, but use the available NIRSpec spectroscopy to remove possible contaminants (see section \ref{sec:bagpipes}). The EGS field was one of the HST/CANDELS fields and thus contains excellent ancillary multi-wavelength observations with the Hubble Space Telescope (HST) and the Spitzer Space telescope as \citep[see, e.g.,][]{Koekemoer2011Candels:Mosaics,Grogin2011Candels:Survey}. This coverage has been further updated with 6 JWST/NIRCam broad bands and one medium band: F115W, F150W, F200W, F277W, F356W, F444W and F410M from the CEERS observations. JWST observations cover an effective area of $\sim34.5$~arcmin$^2$ in multi-band photometry, down to a depth of $28.8$~AB mag in F444W \citep[see also][]{Weibel2024GalaxyObservations}. 
For a more detailed overview of the CEERS survey see \citet{Bagley2022CEERSResults} and \citet{Finkelstein2022AImaging}. 
\begin{figure}
    \centering
    \includegraphics[width=1\linewidth]{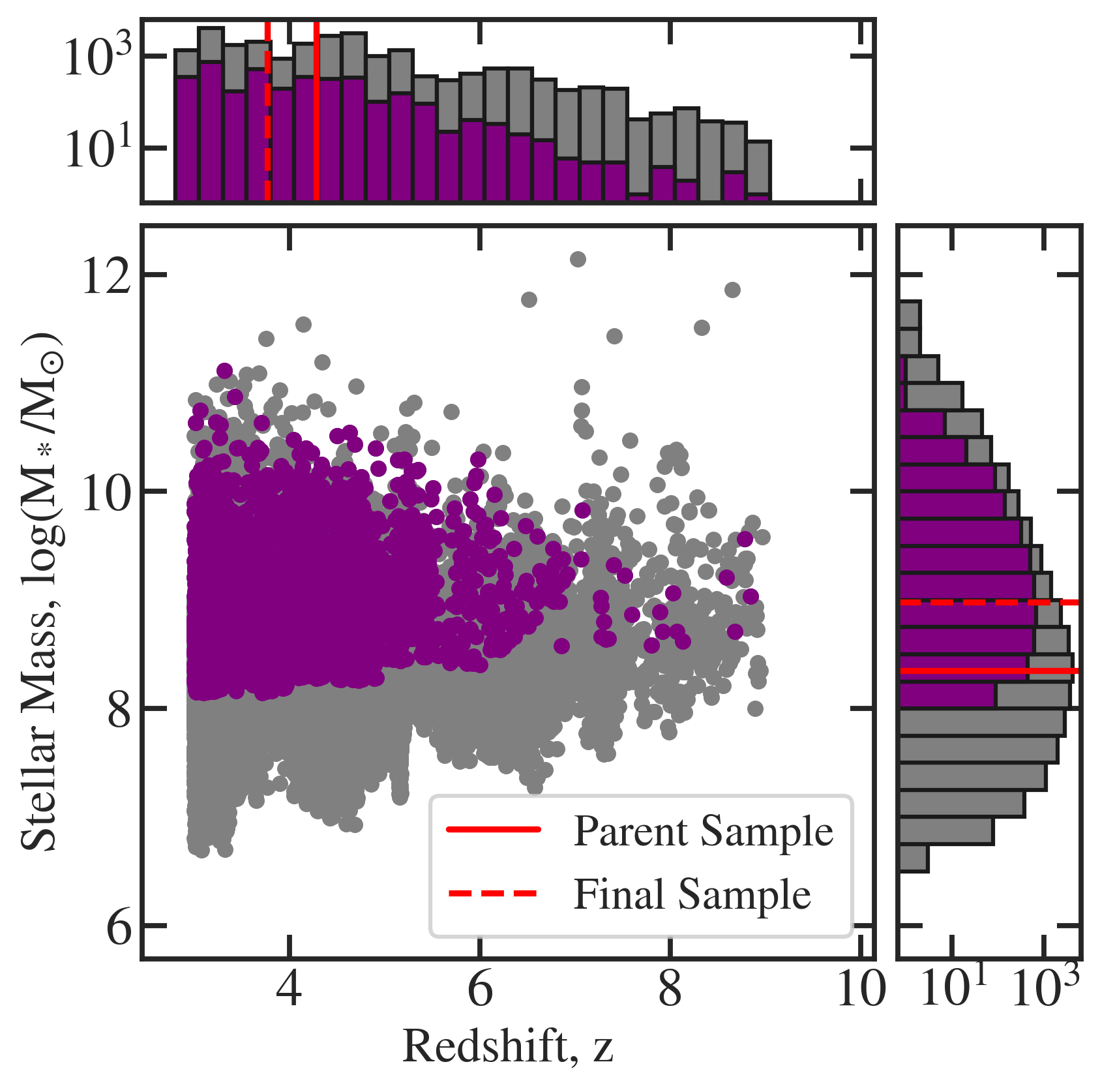}
    \caption{Stellar mass to redshift distribution of the parent (grey points) and final (purple points) sample used in this paper for which reliable multi-wavelength size measurements can be performed (see text for details). The median stellar mass and redshift are shown as red lines.}
    \label{fig:mass_z_dist}
\end{figure}

\subsection{PRIMER-UDS and PRIMER-COSMOS}

Public Release Imaging For Extragalactic Research (PRIMER, Programme ID 1837, PI Dunlop: \citealt{Dunlop2021PRIMER:Research})  is a JWST Cycle 1 general observing program taken in the well-studied CANDELS fields UDS and COSMOS. As part of this program, these fields were covered with JWST/NIRCam imaging in 8 filters: F090W, F115W, F150W, F200W, F277W, F356W, F410M and F444W. The total area of the imaging in both fields is $224.5$~arcmin$^2$ and $127.1$~arcmin$^2$, respectively, with a depth of 28.3 mag in F444W.

\subsection{Photometric Catalogue and Selection of $z\geqslant3$ Sources}
\label{sec:catalogs}

In this work, we use the same multi-wavelength catalogues and galaxy samples in the three fields as described in \citet{Weibel2024GalaxyObservations}. Below, we provide a short overview, but we refer the reader to the previous paper for more details.

Sources are detected from an inverse-variance weighted stack of the long-wavelength data in the F277W, F356W, and F444W filters using the software \texttt{SourceExtractor} \citep{Bertin1996SExtractor:Extraction}. Photometry is then measured in small circular apertures in the PSF-matched images, before being corrected to total fluxes using the Kron apertures and a small correction for remaining flux in the wings of the PSF. Our catalogues include all the available NIRCam filters in these fields as well as the ancillary HST ACS (F435W, F606W, F814W) and WFC3/IR images (F105W, F125W, F140W, F160W). Our catalogues thus include 7 HST as well as 7 or 8 NIRCam filters for the CEERS and the two PRIMER fields, respectively.  

In this paper we aim to measure the sizes studies of galaxies using JWST at over the redshift range of $3\le z < 9$. We select our sample sources based of their photometric redshift based on \texttt{eazy} \citep{Brammer2008EAZY:CODE}, which is run on the photometric catalogues using the \texttt{blue\_sfhz} template set\footnote{\url{https://github.com/gbrammer/eazy-photoz/tree/master/templates/sfhz}}. The parent sample of galaxies is selected using the following criteria: 
\\
\begin{itemize}
    \item[$\bullet$] The best fit photometric redshift ($z_{phot}$) is $> 3$
    \item[$\bullet$] The probability of the best fit photometric redshift, $P(z_{phot} > 2.5)$, is $> 80\%$.
    \item[$\bullet$] The signal-to-noise ratio (SNR) of the source is $> 13$ in the inverse-variance weighted stacked F277W+F356W+F444W detection image.
    \item[$\bullet$] There is available data of the source in all JWST/NIRCam wide bands: F115W, F150W, F200W, F277W, F356W, F410M and F444W.
\end{itemize}
\vspace{0.25cm}

Additionally, we only include sources for which none of the flags described in \citet{Weibel2024GalaxyObservations} are set that identify diffraction spikes of bright stars, residual hot pixels, or other artefacts.

\subsection{Stellar mass and other physical properties}
\label{sec:bagpipes}

We use \texttt{Bagpipes} \citep{Carnall2018InferringMechanisms} to measure the physical properties -- including stellar masses -- and final photometric redshifts of the $z\geqslant3$ sample identified above. Our setup is again identical to the one used in \citet{Weibel2024GalaxyObservations}. Briefly, \texttt{Bagpipes} is run with the following setup: a delayed tau star formation history (SFH) with broad uniform priors in age ranging from 0.01 to 5 Gyr. The allowed range of the ionisation parameter is extended to logU$=$(-4,-1) in order to account for the strong rest-frame optical emission lines that are observed in early galaxies. The \texttt{BPASS-v2.2.1} stellar population models (\citealt{Stanway2018Re-evaluatingPopulations}) are used assuming a broken power-law IMF with slopes of
$\alpha_1 = -1.3$ from 0.1-0.5 $M_{\odot}$ and $\alpha_2 = -2.35$ from 0.5-300 $M_{\odot}$. Finally, a Calzetti dust attenuation curve \citep{Calzetti2000THESTORCHI-BERGMANN} is assumed with a uniform prior on the extinction parameter ${\rm A_V}\in($0, 5) as well as a uniform prior on the stellar metallicity ${\rm Z}\in($0.1, 1$)\,{\rm Z}_\odot$.

The above setup results in satisfactory SED fits for most galaxies, and thus in reliable stellar mass estimates. The photometric redshift estimates are validated based on existing spectroscopic redshifts in \citet{Weibel2024GalaxyObservations}, showing good accuracy with $\langle\Delta z\rangle/(1+z)=0.03$ and a small outlier fraction of $\sim5\%$. In the following, we thus use these redshifts and stellar mass estimates to determine size--mass relations.

\subsection{Sample Selection} 
\label{sec:samp_select}

The $z\geqslant3$ sample from Section \ref{sec:catalogs}, contains both star-forming and quiescent sources. In this paper we aim to fit the size--mass relation of star-forming galaxies and thus need to remove the passive/quiescent sources. Therefore, we use the sSFR estimate from the \texttt{Bagpipes} SED fits. We remove sources with sSFR<$0.2/t_H$, the same limit as used, e.g., in \citet{Carnall2020Timing5}. If this is not taken into account then the size--mass relation flattens in all NIRCam bands, due to quiescent galaxies being more compact (\citealt{VanDerWel20143D-HST+CANDELS:3, Mowla2019COSMOS-DASH:CANDELS/3D-HST,Cutler2022DiagnosingSurvey, Nedkova2021ExtendingCANDELS, Ito2023Size-StellarFields}). The size of the selected star-forming sample is N=23,650 and we present this sample in Fig. \ref{fig:mass_z_dist} as grey points.

To avoid biases in our size--mass relations we use the  80\% mass completeness limits estimated in \citet{Weibel2024GalaxyObservations} per field and redshift bin. Without the mass limits our sample is biased to the bright and compact low-mass sources, resulting in a steeper size--mass relation.

Before proceeding with the derivation of the size--mass relation of galaxies, we have to ensure that the individual measurements are reliable. 
In particular, model fitting is naturally more reliable for sources with higher SNR. While the simultaneous fits across wavelength in principle allow us to relax the SNR requirements somewhat compared to a single-band fit, we still find that measurements become unreliable for sources with SNR$<10$. We find that the bluest bands, that have lower SNRs, are fit with large and unrealistic effective radii (up to 40kpc) if a SNR cut is not applied. Therefore, we require that all our sources are at least a 10$\sigma$ detection in \textit{each} of the six NIRCam filters (excluding F410M) that we use in the fits (see Section \ref{sec:galfitm}). The medium band is excluded as it may be biased to extreme line emission in these high-z star-forming galaxies e.g. O[III]+H$_\beta$. With these SNR limits, our sample will not include any extremely red sources that disappear in shorter wavelength bands, such as HST-dark galaxies \citep{Barrufet2023UnveilingData,Williams2023The8}. For such sources with non rest-UV detections, we can not derive reliable rest-UV sizes. 

One last class of sources we remove are so-called Little Red Dots (LRDs). These have originally been identified as broad-line emitters with point-like morphology in the reddest NIRCam filters \citep[e.g.,][]{Matthee2023LittleSurveys,Labbe2023UNCOVER:ALMA}. While some of these sources can still be dominated by star-formation \citep[e.g.,][]{Williams2023The8,Perez-Gonzalez2024WhatEdition}, spectroscopy has revealed broad emission lines for the majority of these sources, consistent with an AGN component \citep[e.g.,][]{Greene2024UNCOVER5,Wang2024RUBIES:JWST/NIRSpec}. Since we are only interested in estimating sizes for star-forming galaxies in this work, we remove these probable AGNs. To do this, we apply the LRD colour selection from \citet{Labbe2023UNCOVER:ALMA} (removing a total of 126 LRDs and compact red sources; \citealt{Weibel2024GalaxyObservations}) in addition to a size constraint as adopted in \citet{Gottumukkala2024Unveiling8} which removes a further 17 LRDs.

After applying the mass completeness, sSFR and SNR cuts as well as removing the LRDS, we have a parent sample of $3\le z < 9$ star-forming galaxies with size 4116.

\subsection{Point Spread Functions (PSFs)}
In order to measure accurate intrinsic sizes of galaxies, the use of reliable PSFs are crucial. In this work, we tested two different versions of PSFs, 1) created directly from stars in the images using \texttt{ePSF} from the \texttt{Photutils} package and 2) theoretical PSFs created using WebbPSF models \citep{Perrin2014UpdatedWebbPSF} that are drizzled in the same way as the science images. The first approach has the advantage that it is as close to the data as possible. However, in practice, it can be difficult to identify enough stars that are not saturated and contaminated by neighbouring sources. The wings of the PSFs can also be missed if the star cutouts are too small. This can also affect the PSF matched fluxes and thus the physical parameters, e.g. stellar mass measured in SED fitting. 

While the differences between the curves of growth is negligible between the two PSF models, the WebbPSF approach provides higher signal-to-noise models. We tested the two methods by fitting PSF models to stars in the CEERS field and found that the WebbPSF models performed slightly better than our empirical star models, resulting in smaller residuals on average. Therefore, we use the WebbPSF models in the following analysis.

\section{Method}
\label{sec:galfitm}

\begin{figure*}
    \centering
    \includegraphics[width=0.95\textwidth]{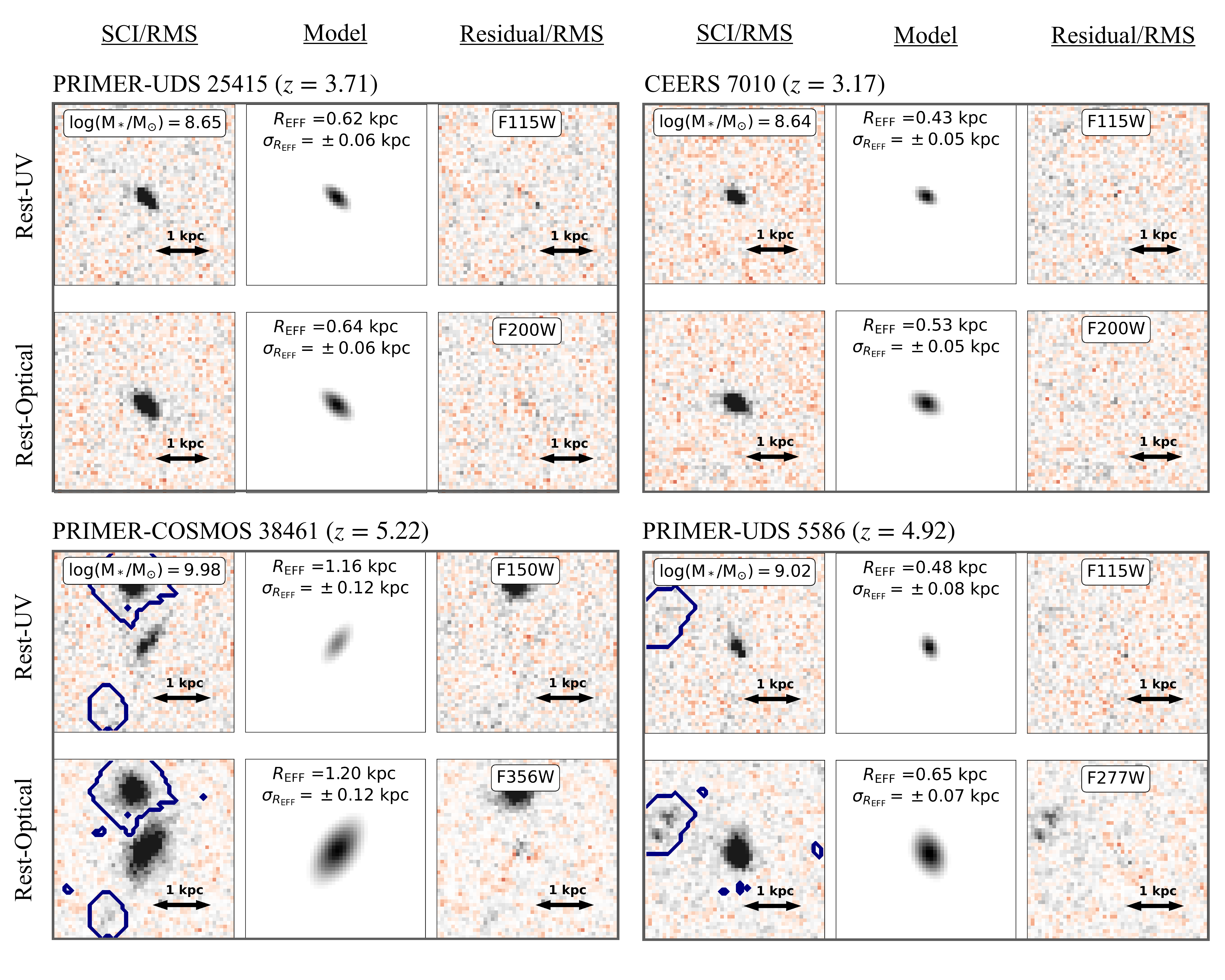}
    \caption{Rest-frame UV and rest-frame optical fits for four star-forming galaxies from our sample (in 2\arcsec\, images). In each block, the left column shows the signal-to-noise ratio images (science/sigma images) with contours showing the mask supplied to \texttt{GalfitM} (see Section \ref{sec:galfitm}). The middle column shows the best fit PSF-convolved \texttt{GalfitM} S\'ersic model to the source. The final box shows the signal-to-noise ratio of the residual image (residual/sigma images). For details on our \texttt{GalfitM} S\'ersic model fitting see Section \ref{sec:galfitm}.}
    \label{fig:example_sources}
\end{figure*}

\subsection{Wavelength Dependent Galaxy Size Measurements}

To derive galaxy sizes, we fit single S\'ersic models with a parametric model fitting tool called \texttt{GalfitM} (\citealt{Hauler2012MegaMorph-multi-wavelengthSurveys, Vika2013MegaMorph-multiwavelengthFar}) to our selected $z>3$ sample. 
\texttt{GalfitM} is an extension of the tool \texttt{Galfit} \citep{Peng2010DETAILEDMODELS}. The latter is a tool that fits the 2D surface brightness of a galaxy, after convolving a user-defined parametric model (e.g. S\'ersic Model) with the PSF of the observation.
\texttt{GalfitM} extends this work by fitting the 2D surface brightness of sources across multiple bands simultaneously. The wavelength dependence of the fitted parameters can be controlled using a polynomial, allowing for varying degrees of freedom between bands (set by a Chebyshev parameter).

\texttt{GalfitM} is run on stamps of 6$\times$6 arcsec centred on each of our galaxies in the sample. Neighbouring sources are masked using a dilated version of the segmentation maps derived from \texttt{SourceExtractor}. This dilation is needed to remove external halo light from bright neighbours. We also derive a local background estimate for each source using a 3-$\sigma$ clipped mean of all unmasked pixels. The background value is supplied to \texttt{GalfitM} and kept fixed while we fit for the S\'ersic parameters for our central sources.

The initial guesses for individual S\'ersic parameters are derived from the \texttt{SourceExtractor} catalogues. This includes the central coordinates ($x,y$), magnitude, axis ratio ($b/a$), and position angle ($\theta $).
The initial value of the effective radius and magnitude is taken to be the semi-major axis and the aperture corrected flux measured by \texttt{SourceExtractor}. The effective radius is defined as the semi-major axis of the best fit ellipse to the galaxy profile. The initial S\'ersic index is set to 1.5, as our sample are expected to be star-forming galaxies and closer to a disk-like structures. During the fits we set some constraints; the S\'ersic values $n$ are allowed to vary between 0.2 to 8, the effective radii to lie between 0.3 to 400 pixels (where 1 pixel = 0.04\arcsec ), the axis ratios between 0.01 to 1 and magnitudes between 0 to 50. 

We perform these S\'ersic fits simultaneously over all NIRCam filters that are available in all our fields, i.e., F115W, F150W, F200W, F277W, F356W, F410M, and F444W. For these fits, we allow the magnitude and effective radii to vary independently across the different bands, in order not to bias our results. However to minimise the number of free parameters, we fix the axis ratio, position angle, and coordinates, because we do not expect these values to change significantly with wavelength.
The S\'ersic index was allowed to vary quadratically with wavelength, as this reduced the number of unrealistic models, while not biasing the results.

Figure \ref{fig:example_sources} shows four examples of input sources with their corresponding \texttt{GalfitM} model in their rest-UV and rest-optical bands. Residuals are also included to show the quality of the S\'ersic fits.

\subsubsection{\texttt{GalfitM} Parameter Uncertainties}
\label{sec:galfit-err-estimates}

\texttt{GalfitM} uncertainties are generally found to be underestimated\footnote{This has been discussed by the author of \texttt{Galfit} here: \url{https://users.obs.carnegiescience.edu/peng/work/galfit/TFAQ.html\#errors} under the topic `Why are the errorbars / uncertainties quoted in "fit.log" often so small?'.}. This occurs because the software assumes that the model is a perfect fit to every unmasked pixel in the image. Therefore, the residual image should only contain Poisson noise \citep{Hauler2012MegaMorph-multi-wavelengthSurveys}. However, this is unrealistic for real astronomical data. 

To estimate realistic uncertainties, we thus follow a similar approach to \citet{VanDerWel2012STRUCTURALCANDELS} and run a Monte-Carlo simulation on simulated images with known input model parameters, but using real blank field images. To do this, we identify 150 blank patches in the CEERS fields of the same size as our cutouts (6\arcsec\, on a side). 
A S\'ersic model, with parameters in the range of our real galaxy sample is added to each of the 150 blank field cutouts. This is done for 250 S\'ersic models which were randomly chosen from our sample.
For each of these simulated cutouts we further derived a corresponding RMS map that includes the Poisson noise of the model galaxy. 

\texttt{GalfitM} was then run on each of the simulated images in the same way as our real data run (as discussed in Section \ref{sec:galfitm}). This includes starting from the same input parameters, constraints and using the same Chebyshev parameters as for the real galaxy fits. The only parameter that was changed was the background estimate, but the method for measuring this estimate is the same method used on the real data.
For each simulated S\'ersic model, we thus obtain 150 fits from the different blank field stamps. The standard deviation between these outputs provides us with an estimate of the real uncertainties in the different parameters, including the size, $\Reff$. We find that the uncertainties derived from \texttt{GalfitM} are typically underestimated, by factors up to $\sim 5\times$. We derive a correction factor through a linear relation between the log of the true standard deviation and the log of the $\Reff$ uncertainty derived by \texttt{GalfitM}. These relations are shown in the appendix in Fig. \ref{fig:simulated-galfitm-errors_std-err}. Throughout the paper, we now use this empirically calibrated correction factors when quoting uncertainties on galaxy sizes.  

\begin{table}[hb]
\centering
\begin{tabular}{ccccc}
\hline
\\[2pt]
$z$ & CEERS & PRIMER & PRIMER  & Total\\
& & -UDS & -COSMOS & \\
\\[-5pt]
\hline \hline
\\[-4pt]
$3\le z <4$ & 678 & 698 & 566 & 1942 \\[3pt]
\hline
\\[-4pt]
$4\le z <5$ & 372 & 478 & 316 & 1166 \\[3pt]
\hline
\\[-4pt]
$5\le z <6$ & 155 & 78 & 92 & 325 \\[3pt]
\hline 
\\[-4pt]
$6\le z <9$ & 47 & 26 & 30 & 125 \\[3pt]
\hline 
\end{tabular}
\caption{Total number of sources in each JWST field for each redshift bin.}
\label{tbl:sample-size}
\end{table}

\begin{figure*}[ht!]
    \centering
    \includegraphics[width=1\linewidth]{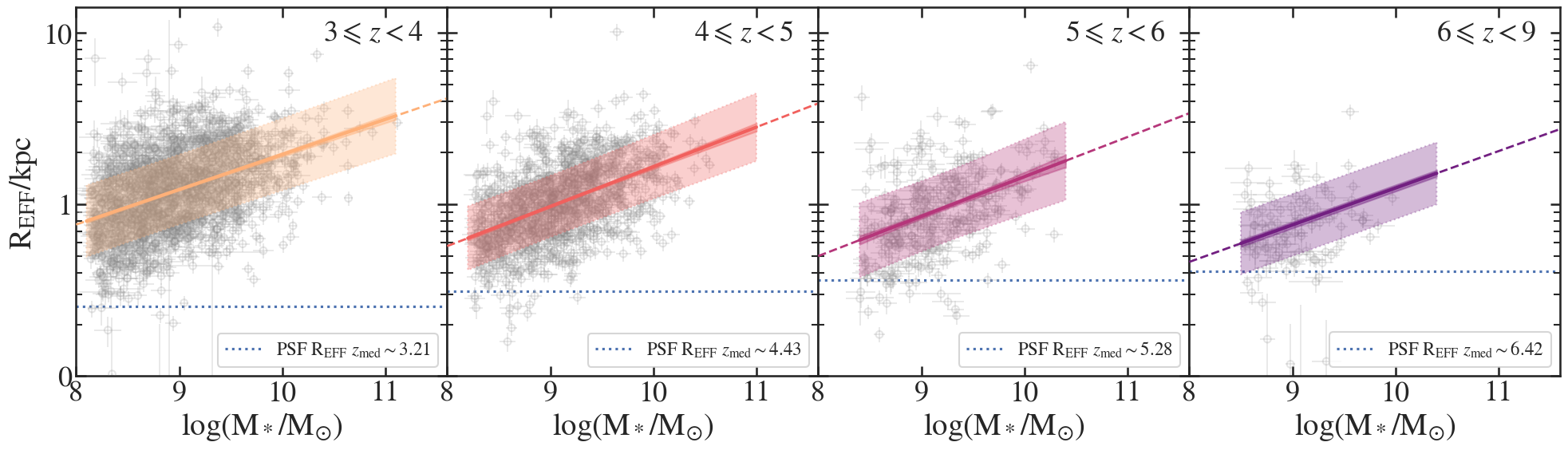}
    \caption{The rest-optical size--mass relation in four redshift bins: $3\le z <4$, $4\le z <5$, $5\le z <6$ and $6\le z <9$. The median fit is shown as a solid black line, the error on the fit is shown as the dark shaded region and the intrinsic scatter is shown in the light grey region. Data points used in the size--mass fitting are shown as grey open circles. The dotted line shows the PSF size at the median redshift of that bin. All points below the PSF $\Reff$ are not included in the size--mass relation fitting. Due to the small number of galaxies in the highest redshift bin, we fix the slope of the size--mass relation in that bin and only fit for the intercept and the intrinsic dispersion (see Section \ref{sec:fitting_size-mass-relation}).}
    \label{fig:rest-opt-sm_allz_w-datapoints}
\end{figure*}

\subsection{Flags and Source Removal} 

Even after our initial SNR cuts, a small number of galaxies cannot be fit well with \texttt{GalfitM}. Following the practice in previous works, we thus removed sources for which the model fits reach the limits of our constraints. We thus discarded sources with S\'ersic index of $>7.8$ or $<0.25$, axis ratios of $<0.02$ or $>0.95$, and effective radii that were fit to be larger than the cutout size. 

Finally, we visually inspected all the remaining fits and their residuals. Sources are removed if they have extremely bright and large neighbours within a few pixels from their segmentation map. Flux from such nearby neighbours is not completely removed by the applied masks, creating a bright background and affecting the fitting.  
We also check that individual fits do not have strong residuals following the residual flux fraction (RFF) method used in \citet{Ward2023EvolutionCEERS} and \citet{Ormerod2023EPOCHSObservations} but find that most of our fits lied below the 0.5 cut used in these works. 

After removing all these sources from our parent sample, we are left with a final sample of 3529 star-forming galaxies. The breakdown by field can be seen in Table \ref{tbl:sample-size} and the distribution of redshift and stellar mass for both the parent and final sample are shown in Fig. \ref{fig:mass_z_dist}.

\subsection{Measuring the Size--mass Relation}
\label{sec:fitting_size-mass-relation}
The galaxy size--mass relation has been shown to be well-approximated by a simple power law. Following past literature, we thus model the size--mass relation in the same way. Specifically, we use the hierarchical Bayesian linear regression tool \texttt{Linmix}\footnote{ \url{https://github.com/jmeyers314/linmix}} \citep{Kelly2007} to fit the mean relation as well as an intrinsic dispersion around the mean, $\sigma_{\log \Reff}$. We fit a log-normal distribution to the size--mass relation as shown in Equation \ref{equ:size_mass_linmix}.

\begin{equation}
         \log \left( \frac{\Reff}{\text{kpc}} \right) \sim  \alpha \log \left( \frac{M_*}{10^9 M_{\odot}} \right) + \log \left( \frac{A}{\text{kpc}} \right) +  \mathcal{N} (1, \sigma^2_{\log \Reff})
         \label{equ:size_mass_linmix}
\end{equation}

Where $\Reff$ is the effective radius, $\alpha $ is the slope of the relation, $\rm M_*$ is the stellar mass of a galaxy in solar masses and $\rm \log A$ is the intercept, i.e., the mean effective radius of galaxies of stellar mass $\rm 10^{9}M_{\odot}$. Finally, $\sigma_{\log \Reff}$ is the intrinsic scatter, which measures the distribution in galaxy sizes at fixed stellar mass. 

\texttt{Linmix} allows us to account for the uncertainties in both stellar mass and size measurements and it provides us with posterior samples for all the parameters from which we can derive the best-fit values as well as the uncertainties. We use \texttt{Linmix} to fit for $\alpha$, $\rm \log A$ and $\log \sigma_{\log \Reff} $ for each NIRCam band in the three fields, in four redshift bins: $3\le z<4$, $4\le z<5$, $5\le  z<6$ and $6\le z<9$. We exclude sources with $\Reff$ smaller than the PSF $\Reff$ at the sources photometric redshift. Due to low-number statistics in the $6\le z <9$ sample, we can not constrain the size--mass relation without fixing the slope of the relation. Thus, based off our rest-optical parameter evolution measurements in Section \ref{sec:discussion_rest-opt_size-evolution} we have fixed the slope of our highest redshift bin to 0.21. Further work is needed to constrain the $z>6$ size--mass relation and can be achieved by combining deep data sets, e.g. NGDEEP and JADES. The results of our fits are discussed in Section \ref{sec:results}.

\begin{figure}[tbp]
    \centering
    \includegraphics[width=1\linewidth]{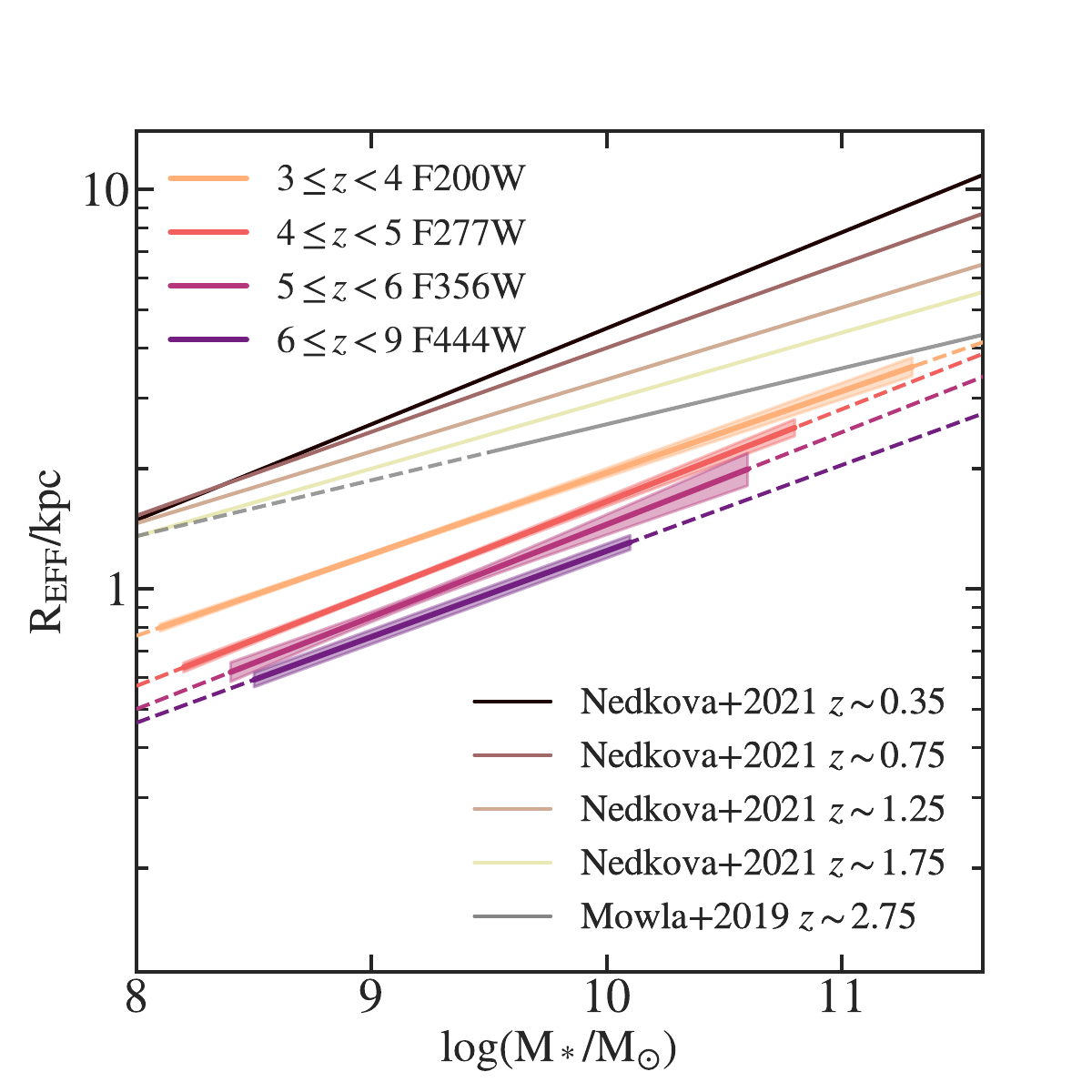}
    \caption{Comparing our JWST-based $z>3$ rest-optical size--mass relations to low-$z$ HST star-forming size--mass relations from \citet{Nedkova2021ExtendingCANDELS} and the $z=2.75$ $\lms > 10$ star-forming relation from \citet{Mowla2019COSMOS-DASH:CANDELS/3D-HST}. The $1\sigma$ error on the fits are shown as the shaded regions around the main fit. The intrinsic scatter is not shown in these plots for better visualisation. The slope of the $6\le z <9$ relation has been fixed for reasons discussed in Section \ref{sec:fitting_size-mass-relation}.}
    \label{fig:sm_rest-opt_evolution_w-hst-sm}
\end{figure}

\begin{figure*}
    \centering
    \includegraphics[width=1\textwidth]{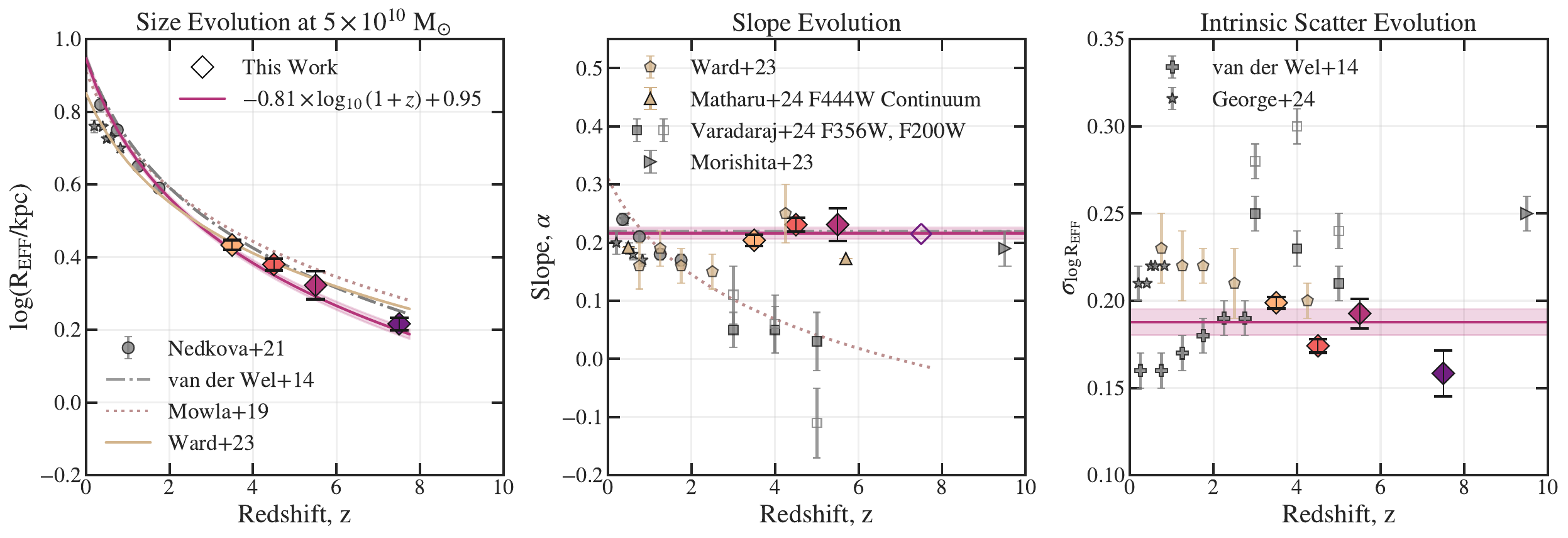}
    \caption{Redshift evolution in the best-fit star-forming size--mass relation parameters. The left panel shows the size evolution for galaxies at a fixed stellar mass of $M_* = 5\times 10^{10} M_{\odot}$ from $0.35\le z<9$. Centre panel shows the evolution of the slope, $\alpha$ and the right panel shows the evolution in the intrinsic scatter, $\sigma_{\log \Reff}$. Our rest-optical measurements are shown as solid diamonds and their measured evolution function in rest-optical is plotted as solid line in each panel. Low-$z$ measurements from HST-based studies (\citealt{Nedkova2021ExtendingCANDELS}, \citealt{Mowla2019COSMOS-DASH:CANDELS/3D-HST}, \citealt{George2024TwoNewcomers} and \citealt{VanDerWel2012STRUCTURALCANDELS}) are shown in grey. JWST-based studies from \citet{Ward2023EvolutionCEERS} and \citet{Matharu2024ASpectroscopy} are shown in tan. The JWST-based work based off UV-bright ground based selected star-forming galaxies from \citet{Varadaraj2024ThePRIMER} are shown in grey squares. The HST-based parameterised functions fit to the redshift evolution from  \citet{Mowla2019COSMOS-DASH:CANDELS/3D-HST} and \citet{VanDerWel2012STRUCTURALCANDELS} as well as the JWST-based size-evolution from \citet{Ward2023EvolutionCEERS} are also included. See Table \ref{tbl:sm_linear_params_allz_allbands} for values. In the central panel, the slope of our $6<z<9$ bin is shown as an open diamond as it was fixed during the size--mass fitting (see Section \ref{sec:fitting_size-mass-relation}).}
    \label{fig:linear-params-evolution_hst-comp}
\end{figure*}

\section{Results}
\label{sec:results}
Using the sizes derived in Section \ref{sec:galfitm}, we fit the size--mass relation (Equation \ref{equ:size_mass_linmix}) of star-forming galaxies in four redshift bins: $3\le z < 4$, $4\le z < 5$, $5\le z < 6$ and $6\le z < 9$. 

With the large wavelength coverage of the JWST/NIRCam imaging conducted in all the fields; CEERS, PRIMER-UDS, and PRIMER-COSMOS, we can study the sizes of galaxies in 7 NIRCam bands (F115W, F150W, F200W, F277W, F356W, F410M and F444W). In the following, we will first discuss the evolution of rest-frame optical size--mass relations, before we compare the size estimates at different wavelengths. 
For the discussion in rest-UV ($2000-2500$\AA) we use F115W, F150W, F200W and F277W, for each redshift bin respectively. Similarly for rest-optical ($4500-5500$\AA), we use F200W, F277W, F356W and F444W.

When presenting the size--mass relation we plot two shaded regions around the median fit. The first shaded region is darker and represents the $1\sigma$ uncertainty on the mean relation, while the second shaded region is lighter and shows the intrinsic scatter. In some of the plots presented in this paper the intrinsic scatter has been removed for clarity.

\subsection{Rest-Frame Optical Size Evolution}
\label{sec:result_rest-opt_size-evolution}

Now with the high sensitivity and large near-infrared wavelength coverage of JWST, we can finally extend rest-optical galaxy size measurements to $z\sim 9$ and down to stellar masses of $\lms \sim 8-9$. 

We present our rest-optical star-forming size-mass relations from $z=3$ to $z=9$ in Fig. \ref{fig:rest-opt-sm_allz_w-datapoints} and Fig. \ref{fig:sm_rest-opt_evolution_w-hst-sm}.  Our $6\le z < 9$ redshift bin suffers from low number statistics, as shown by the data points in the right panel of Fig. \ref{fig:rest-opt-sm_allz_w-datapoints}. Therefore, we fix the slope of the size--mass relation to 0.215, which is the average slope calculated in Section \ref{sec:results_rest-opt_slope-evolution}. 
In Fig. \ref{fig:sm_rest-opt_evolution_w-hst-sm}, we find that the rest-optical size--mass relation evolves over the redshift range $3\le z<9$, with galaxies becoming smaller and more compact at higher redshifts. 

It is easier to understand the evolution of the rest-optical, size--mass relation by studying the parameters and their evolution with redshift.
Thus, we have parameterised the redshift evolution of the best-fit rest-optical size--mass parameters at different epochs between $z=3$ and $z=9$. The evolution of these parameters are presented in Fig. \ref{fig:linear-params-evolution_hst-comp}. We present the results of each parameter in Section \ref{sec:results_rest-opt_5x10-evolution}, Section \ref{sec:results_rest-opt_intrsigma-evolution} and Section \ref{sec:results_rest-opt_slope-evolution}.

Also in this section, we compare our rest-frame optical size--mass relations to low-z HST studies from \citet{Nedkova2021ExtendingCANDELS}, \citet{Mowla2019COSMOS-DASH:CANDELS/3D-HST} and \citet{VanDerWel20143D-HST+CANDELS:3} as well as recent JWST-based studies from \citet{Ward2023EvolutionCEERS}, \citet{Matharu2024ASpectroscopy}, \citet{Morishita2023Enhanced14} and \citet{Varadaraj2024ThePRIMER}.
Each of these works measured the size--mass relation over different stellar mass ranges and there are systematics due to different methodologies, thus it is important to highlight these differences before comparing.

\citet{Nedkova2021ExtendingCANDELS}, \citet{VanDerWel20143D-HST+CANDELS:3}, and \citet{Mowla2019COSMOS-DASH:CANDELS/3D-HST} used HST data, applying UVJ color selections to separate quiescent and star-forming galaxies. \citet{Nedkova2021ExtendingCANDELS} measured rest-5000\AA\, sizes of $0<z<2$ star-forming galaxies ($8<\lms<11.5$), while \citet{VanDerWel20143D-HST+CANDELS:3} covered $0<z<3$ galaxies ($9<\lms<11.5$) and \citet{Mowla2019COSMOS-DASH:CANDELS/3D-HST} extended this to include $ \lms > 11.3$ galaxies. Both \citet{VanDerWel20143D-HST+CANDELS:3} and \citet{Mowla2019COSMOS-DASH:CANDELS/3D-HST} used single S'ersic models in the $H_{160}$ band with \texttt{Galfit}. \citet{Nedkova2021ExtendingCANDELS} used \texttt{GalfitM} and the Chebyshev polynomials to calculate then rest-5000\AA\, sizes.
With the recent JWST data, \citet{Ward2023EvolutionCEERS} studied $\lms > 9.5$ galaxies ($0.5<z<5.5$) selected with SED fitting from the CEERS dataset. This work is similar to our approach in this paper. \citet{Matharu2024ASpectroscopy} differs in that their measurements are made on stacks of hundreds of galaxies in the F444W band from FRESCO ($8<\lms<10$), while \citet{Varadaraj2024ThePRIMER} focused on bright LBGs ($3\le z<5$, $9<\lms<11$) selected by ground-based data but observed in the PRIMER dataset. Finally, \citet{Morishita2023Enhanced14} and \citet{Ono2023CensusFormation} measured sizes of $4<z<14$ galaxies, with \citet{Morishita2023Enhanced14} using a combined dataset from nine JWST surveys and \citet{Ono2023CensusFormation} selecting galaxies from CEERS.

\begin{figure*}[h!]
    \centering
    \includegraphics[width=1\textwidth]{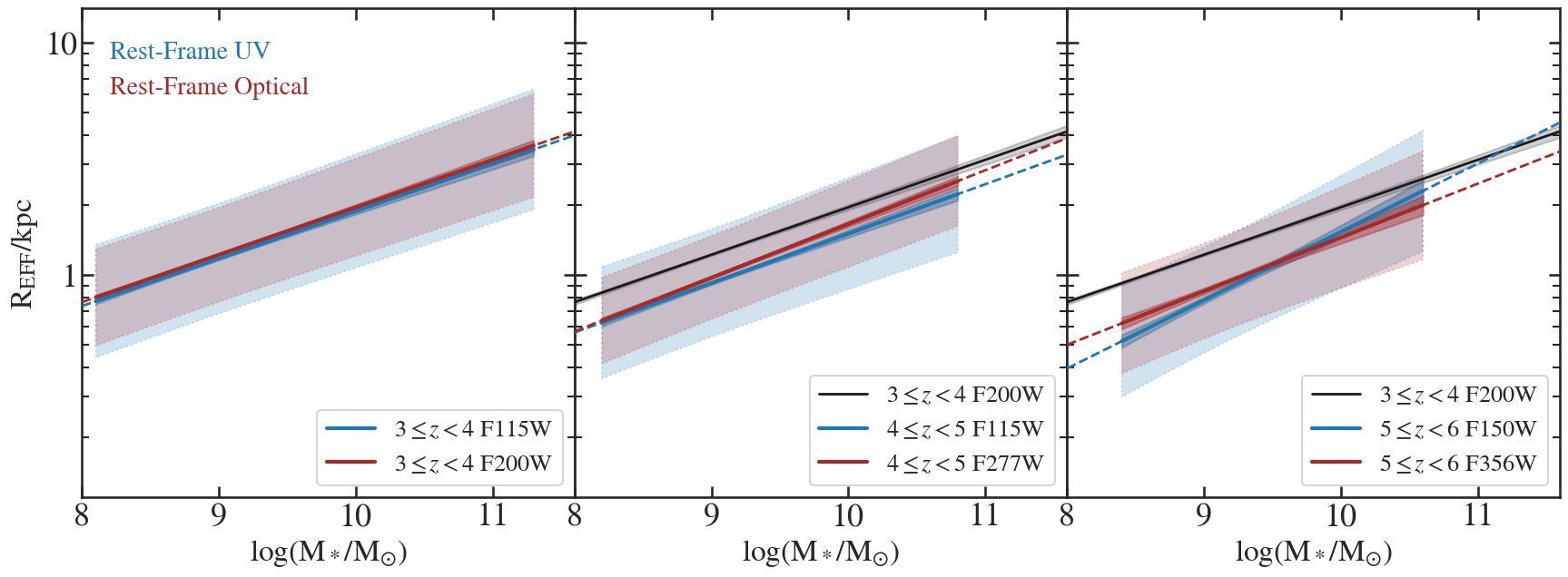}
    \caption{The star-forming size--mass relations in bands corresponding to rest-UV (blue) and rest-optical (red) in three redshift bins (from left to right): $3\le z < 4$, $4\le z < 5$, and $5\le z < 6$. The median size--mass fit is surrounded by two shaded regions: the first shaded region is the $1\sigma$ error on the fit and the second is the median intrinsic scatter. The star-forming $3\le z < 4$ size--mass relation is plotted as a black dashed line in three of the panels. The rest-UV and rest-optical $z<5$ relations are within $1\sigma$ of each other but this changes at $z>5$ where star-forming galaxies at $\lms < 9$ are $>2\sigma$ smaller in rest-UV.}
    \label{fig:sm_allz_rest-uv_rest-opt}
\end{figure*}

\subsubsection{Redshift Evolution of Star-forming Galaxies of Stellar Mass $5\times 10^{10} \rm M_{\odot}$ Rest-optical Sizes}
\label{sec:results_rest-opt_5x10-evolution}
The effective radius measurements for galaxies of stellar mass $\rm M_* = 5\times 10^{10} M_{\odot}$ are plotted in the left panel of Fig. \ref{fig:linear-params-evolution_hst-comp}. This stellar mass is chosen such that we can fairly compare our measurements with previous estimates from the literature. We note, however, that our samples of galaxies at those high masses is small and the values we show mostly rely on the size--mass relation fits to lower mass galaxies.

From the plot it is evident that the size evolution in star-forming galaxies at this fixed mass is well fit by a log-linear relation over the period of $0<z<9$:

\begin{equation}
    \log \Reff = {\beta _z}\times \log_{10}(1+z) +\rm B_z
    \label{equ:R_evolution_z}
\end{equation}

We thus combine our rest-optical size estimates at $z>3$ with values measured at $z<3$ from \citet{VanDerWel20143D-HST+CANDELS:3}, to fit for the parameters in Equation \ref{equ:R_evolution_z} over the total redshift range of $0<z<9$. Over this period, we find that the size-evolution of star-forming galaxies of mass $\rm M_* = 5\times 10^{10}\,M_\odot$ follows the function $\log \Reff = (-0.807\pm 0.026) \times \log_{10}(1+z) +(0.947\pm0.014)$. 
This is shown in the left panel of Fig. \ref{fig:linear-params-evolution_hst-comp} as a solid line and indicates that a typical star-forming galaxy of $\rm M_* = 5\times 10^{10} M_{\odot}$ grows rapidly by a factor of 2$\times$ from $z\sim8$ to $z\sim3$ and then another factor of 3$\times$ to $z\sim0$.

Interestingly, our measured evolution is in relatively good agreement with extrapolations from lower redshifts from \citet{VanDerWel20143D-HST+CANDELS:3} and \citet{Mowla2019COSMOS-DASH:CANDELS/3D-HST}, as well as with recent JWST-based estimates from \citet{Ward2023EvolutionCEERS}. 
Our measured average sizes at $z\sim4-6$, are withing $1\sigma$ of the parameterised evolution predicted by HST-based, $z<3$ studies.

\subsubsection{Evolution of the Distribution in Rest-Optical Sizes of Star-forming Galaxies at $3\le z <9$}
\label{sec:results_rest-opt_intrsigma-evolution}
To study the distribution of galaxy sizes over time, we have plotted our measured intrinsic dispersion values between $3\le z<9$ in the right panel of Fig. \ref{fig:linear-params-evolution_hst-comp}.
While the measured values vary somewhat, we find that the intrinsic dispersions are consistent with a constant value of $\sigma_{\log \Reff} = 0.188 \pm  0.007$, over the redshift range $z=3$ to $z=9$. This is within a $1\sigma $ agreement with the rest-optical intrinsic scatter measured by \citet{VanDerWel20143D-HST+CANDELS:3} at lower redshifts. Our rest-optical intrinsic scatter data points at $z=3.5$, $z=4.5$ and $z=5.5$, are within $1-2\sigma $ of the $z=4.2$ rest-optical and $z=5.7$ stacked F444W intrinsic scatter measurements by \citet{Ward2023EvolutionCEERS} and \citet{Matharu2024ASpectroscopy}, respectively. However, we find $>2\sigma $ differences with the intrinsic scatter measured by \citet{Morishita2023Enhanced14} and \citet{Varadaraj2024ThePRIMER}, who find larger values for galaxies between $4<z<14$ and $3<z<6$, respectively. This difference could be due to both studies fitting the size--mass relation for Lyman break galaxies (LBGs). Also, as discussed in Section \ref{sec:results_sizes_wavedep}, the size of a galaxy varies depending on the wavelength it was measured in. This could also lead to the difference between our work and  \citet{Varadaraj2024ThePRIMER}, who only measures galaxy sizes in F200W and F356w=W. 
 
Except for these two estimates, the intrinsic dispersion measurements of all literature results lie in the range  $\sigma_{\log \Reff} = 0.15-0.25$ dex. 
Overall, these results thus indicate, at fixed stellar mass, that the dispersion of star-forming galaxy sizes around the mean stays rather constant across the full redshift range probed to date.

\subsubsection{Evolution of the Rest-optical Size--Mass Slope}
\label{sec:results_rest-opt_slope-evolution}

Finally, we discuss the slope measurements of the rest-optical size--mass relation, which are plotted in the central panel of Fig. \ref{fig:linear-params-evolution_hst-comp}.
In our sample of star-forming galaxies at $z=3-9$, we find slopes that are consistent with each other within $1-2\sigma$ at all redshift. The average value is found to be $\alpha=0.215\pm 0.009$ over the redshift range of $3\le z<9$. 
This value is consistent within $1\sigma$ with \citet{VanDerWel20143D-HST+CANDELS:3} who found that the average size--mass slope is $\alpha=0.22$, for star-forming galaxies between $0<z<3$. Similar values are also found by \citet{Ward2023EvolutionCEERS} and \citet{Matharu2024ASpectroscopy} at $z\sim4-5$, and even by \citet{Morishita2023Enhanced14} who analysed galaxies out to $z\sim14$.
This could indicate that the steepness of the size--mass relation of star-forming galaxies remains unchanged over the full cosmic history at a value of $\alpha=d\log \Reff / d\log\rm M_*\sim0.2$. 

This result is different from previous extrapolations of lower-redshift trends from, e.g., \citet{Mowla2019COSMOS-DASH:CANDELS/3D-HST}, who found a trend to flatter slopes to higher redshifts (see central of panel Fig. \ref{fig:linear-params-evolution_hst-comp}), in line with the estimates from \citet{Varadaraj2024ThePRIMER} at $z=3-5$. It will thus be important to revisit the size--mass relation estimates in the future with larger samples also extending to lower and higher masses to test whether the slopes really stay constant over redshift, of if they might evolve.

\subsection{The Rest-Optical and Rest-UV Size--Mass Relations Star-Forming Galaxies}
To understand the build up of stellar mass in $z>3$ star-forming galaxies, we investigate galaxy size measurements in rest-UV and rest-optical. The rest-UV and rest-optical size--mass relations in three redshift bins are presented in Fig. \ref{fig:sm_allz_rest-uv_rest-opt}. The highest redshift bin is not shown due to the small number of galaxies that do not allow us to accurately fit for the slope of the size--mass relations (see Fig. \ref{fig:rest-opt-sm_allz_w-datapoints}).

In all three panels we can see that over the stellar mass range of $8 < \lms < 11$, the rest-optical and rest-UV relations are within 1$\sigma$ of each other, except in the $z\sim 5-6$ bin, where the rest-UV size--mass relation appears to be steeper than the rest-optical one. The mean size of $5\le z < 6$ star-forming galaxies with stellar mass $\lms \le 9$, are $> 2\sigma$ smaller in rest-UV than in rest-optical wavelengths. More details are discussed in Section \ref{sec:discussion-multilambda_sizes}. 

Interestingly, at all redshifts, we find the intrinsic dispersion of the size--mass relation to be larger in the rest-UV compared to the rest-frame optical, hinting at a wavelength dependence. We investigate these trends further in the next section.

\subsection{The Wavelength Dependence of the Size--Mass Relation Parameters}
\label{sec:results_sizes_wavedep}
\begin{figure*}[h!]
    \centering
    \includegraphics[width=1\textwidth]{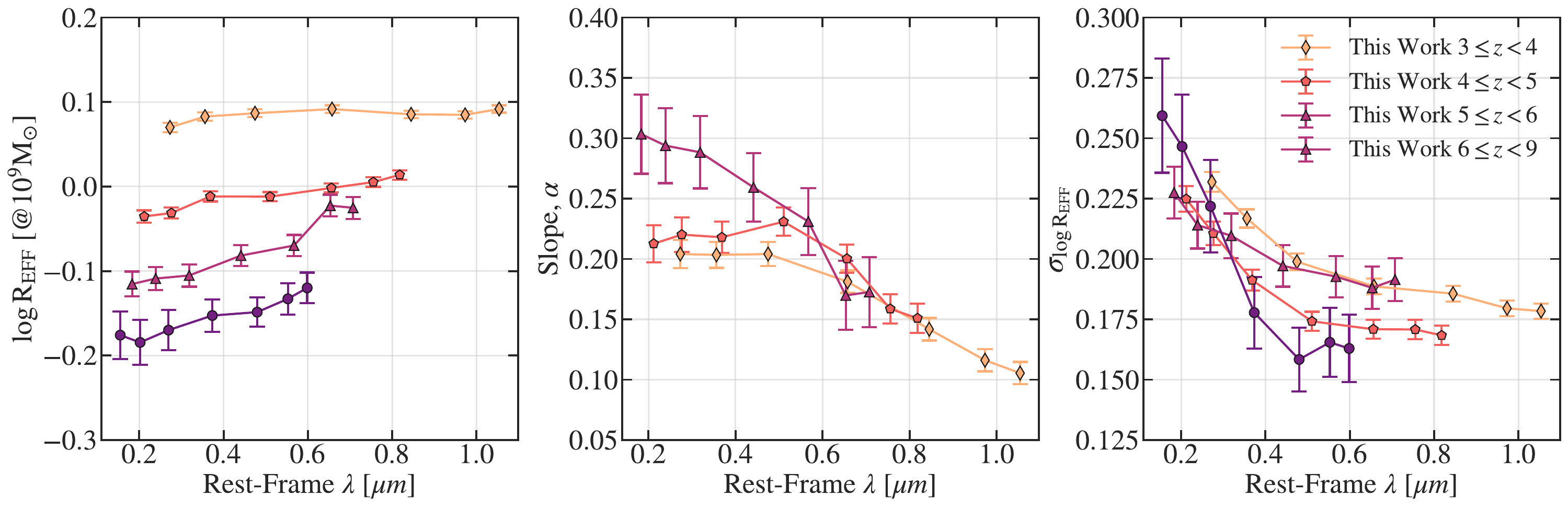}
    \caption{Our best-fit star-forming size--mass relationship parameters: intercept in kpc at $10^9M_\odot$ (left), slope $\alpha$ (centre), and intrinsic scatter $\sigma_{\log  \Reff}$ (right) as a function of rest-frame wavelength (using the median redshift in each redshift bin - see Table \ref{tbl:sm_linear_params_allz_allbands}). The $6\le z<9$ size--mass slope is removed from the central panel because it was fixed with wavelength in the size--mass fitting (see Section \ref{sec:fitting_size-mass-relation}). The trend of a decreasing slope with increasing wavelength at all redshifts is consistent with colour gradients. The large  $\sigma_{\log  \Reff}$ at $\rm \lambda < 0.5\,\mu m$ could be a result of off-centre bursty star-formation in star-forming galaxies at $z>3$.}
    \label{fig:sm_parameters_allz_rest-lambda}
\end{figure*}

Due to the extended NIRCam wavelength coverage, we can investigate the size--mass relation over multiple wavelengths spanning rest-UV and rest-optical, including longer wavelengths than rest-frame $0.5\rm \,\mu m$, (in the three lower redshift bins). This is important to understand the stellar mass build up of galaxies. Fig. \ref{fig:sm_parameters_allz_rest-lambda} shows the variation in the size--mass relation parameters that we fit for as a function of rest-frame wavelength. The rest-frame wavelength for each NIRCam band is calculated using the median redshift of a given bin. These parameters and median redshifts are specified in Table \ref{tbl:sm_linear_params_allz_allbands}.

The left panel of Fig. \ref{fig:sm_parameters_allz_rest-lambda} shows the sizes of $\lms = 9$ star-forming galaxies in four redshift bins. 
In all redshift ranges, the sizes of star-forming galaxies are consistent within $1\sigma$ between the wavelength range of $\rm 0.15\, \mu m < \lambda < 0.5\,\mu m$. This behaviour also continues to $\rm 1\, \mu m$ for $3\le z<5$ star-forming galaxies.
However, there is an indication that the sizes of $\lms = 9$ star-forming galaxies, at $z>5$ positively increase at $\rm \lambda > 0.5\, \mu m$, although this trend is uncertain due to the large errors. 

This figure also shows that for a given rest-frame wavelength, the average size of $\lms=9$ star-forming galaxies decreases to higher redshift. The rate of this decrease is somewhat wavelength dependent, with rest-UV sizes decreasing slightly faster than rest-optical sizes.

To further examine the dependence of galaxy size with stellar mass, we show the slope of the size--mass relation as a function of rest-frame wavelength in the central panel of Fig. \ref{fig:sm_parameters_allz_rest-lambda}, except for the highest redshift bin (for which the slope cannot be measured accurately). For all three redshift bins between $\rm 0.2\, \mu m<\lambda<0.8\,\mu m$, there is a general indication of a decreasing trend with increasing wavelength. I.e., the longer wavelength size--mass relations become flatter. However, given the errors, this downward trend is uncertain.

Focusing on the slopes from the $3\le z< 4$ size--mass relation that we can determine best, we see roughly a constant slope of $\alpha=0.2$ at wavelengths of $\lambda <\rm 0.5\,\mu m$, which then starts to flatten to $\alpha=0.1$ at $\rm \lambda =1\,\mu m$. This trend is also observed in the $4\le z<5$ redshift bin, but is uncertain due to the large errors. We discuss the implications of this on galaxy stellar mass build-up further in Section \ref{sec:dis_z34_colgrads}.

Finally, the right panel in Fig. \ref{fig:sm_parameters_allz_rest-lambda} shows the intrinsic scatter as a function of rest-frame wavelength. For all redshift bins the intrinsic scatter starts high around a value of $\sim 0.22-0.25$ dex in rest-UV and then falls and plateaus to values between $\sim 0.17- 0.19$ dex, at wavelengths $\lambda > \rm 0.5\, \mu m$. This shows that the dispersion in galaxy sizes at a fixed mass become $10\%$ smaller when measured in redder wavelengths compared to rest-UV and this trend is independent of redshifts between $3\le z < 9$, within current uncertainties. 

Overall, our analysis thus reveals that the size--mass relation of star-forming galaxies at $z\sim3-8$ has a higher dispersion and steeper slope in the rest-frame UV compared to longer rest-frame wavelengths. The implications of this are discussed in section \ref{sec:discussion-multilambda_sizes}.

\section{Discussion}
\label{sec:discussion}
In this section, we discuss the two main results presented in this paper: the redshift evolution of the rest-frame optical sizes (Section \ref{sec:discussion_rest-opt_size-evolution}), 
and the wavelength dependence of the size--mass relation parameters (Section \ref{sec:sm_lambda_params_dis}).
We also comment on comparisons with simulations  that have measured the size mass relation at $z>3$ (Section \ref{sec:sim_comp_dis}) as well as comparisons to low-z rest-NIR studies (Section \ref{sec:dis_z34_colgrads}).

\subsection{Analysis of the Rest-Frame Optical Size Evolution}
\label{sec:discussion_rest-opt_size-evolution}

In Section \ref{sec:result_rest-opt_size-evolution} we found that the rest-optical size--mass slope and intrinsic dispersion are consistent with each other over $3\le z<9$ and their evolution can be well described by a constant.
This provides some interesting clues on the relation between the galaxies and their dark matter halos.

Simple galaxy disk models predict a correlation between the dispersion in dark matter halo spin and the amount of scatter in the size-mass relation (see, e.g., \citealt{Mo1998TheDiscs}, \citealt{Ward2023EvolutionCEERS}, \citealt{VanDerWel20143D-HST+CANDELS:3}, and citations within). 
Most simulations have measured the dispersion of the DM spin parameter to be $0.2-0.25$ dex (log base 10) (e.g. \citealt{Maccio2008ConcentrationResults, Burkert2016THE}). This is consistent with our average intrinsic scatter of  $\sigma_{\log \Reff} = 0.188 \pm  0.007$, over the redshift range of $3\le z <9$. This provides further support that the sizes of star-forming galaxies at $z>3$ are indeed driven by their dark matter halo's spins, even in the early Universe.

Additionally, the slope of the size--mass relation carries information on the ratio between stellar and halo masses. As outlined in the introduction,  we expect a slope of $\alpha_\mathrm{DM}=1/3$ ($\rm R_{ virial} \propto M_{\rm halo}^{1/3}$) for DM halos. This is steeper than the constant slope of the size--galaxy mass relationship we found of $\alpha = 0.215\pm 0.009$ at $z=3-9$. Such a difference in slopes can be explained if the relation between the stellar mass and halo mass is not a constant, which has also been found in previous works (e.g., \citet{VanDerWel20143D-HST+CANDELS:3}, \citet{Ward2023EvolutionCEERS} and \citet{Shen2003TheSurvey}).

\subsection{Sizes of Star-forming Galaxies in Multi-Wavelength Observations}
\label{sec:discussion-multilambda_sizes}

\label{sec:sm_lambda_params_dis}
In this section we discuss the results of the wavelength dependence on the size--mass relation and its parameters that we fit for with rest-frame wavelength and redshift.

In the right panel of Fig. \ref{fig:sm_parameters_allz_rest-lambda} we find that the intrinsic scatter, for all redshift bins and over the stellar mass range for which we measure size--mass relations, follows a decreasing trend with increasing wavelength. A larger rest-UV intrinsic scatter, in comaprison to rest-optical, is also seen for $z=5$ bright LBGs studied in \citet{Varadaraj2024ThePRIMER} as well as for star-forming galaxies at $z<1$ studied by \citet{George2024TwoNewcomers}. High levels of intrinsic scatter indicate significant variations in measured quantities. Bursts of star-formation tend to be short lived, compact and not necessarily spatially coherent from galaxy-to-galaxy \citep[see, e.g.,][]{Gimenez-Arteaga2024OutshiningNIRCam}. Recent studies have also shown that $z>4$ star-forming galaxies have bursty star-formation (\citealt{Looser2023JADES:Universe, Carnall2023AHistories, Curti2023JADES:Spectroscopy}). Since rest-UV wavelengths trace star-formation on short timescales ($\sim 100\rm \,Myrs$) significant variations in the rest-UV sizes (plus smaller variations in rest-optical sizes) can therefore be most likely attributed to bursty star formation. 

In the left and centre panels of Fig. \ref{fig:sm_parameters_allz_rest-lambda} we find over the wavelength range of $\rm 0.15\, \mu m<\lambda < 0.6\,\mu m$ no significant wavelength dependence on the intercept and slope. This is also shown in Fig. \ref{fig:sm_allz_rest-uv_rest-opt} for the whole size--mass distribution. this suggests that on average rest-UV and rest-optical sizes are similar for star forming galaxies from $z=3$ to $z=9$, over the mass range we fit for. JWST based studies from \citet{Morishita2023Enhanced14} and \citet{Ono2023CensusFormation} have also found that the ratio between rest-UV and rest-optical sizes are $\sim 1$. \citet{Ormerod2023EPOCHSObservations} also finds that $3<z<8$ star-forming galaxies at $\lms <9.5$ are consistent within $1\sigma$ over the observed wavelength range of $\rm 1\,\mu m < \lambda_{obs} < 5\,\mu m$. Similar sizes, at these low masses, indicate either that these galaxies are young and haven't built up their older stellar population \citep{Morishita2023Enhanced14, Ono2023CensusFormation}, possible outshining effects from highly star-forming clumps \citep{Gimenez-Arteaga2024OutshiningNIRCam}, or compact dust attenuation is biasing the sizes in these wavelengths. 

At wavelengths $\lambda > \rm 0.5\, \mu m$, there is a general decrease in the slope of the size-mass relation. This flattening of the slope is seen up to $\rm 1\,\mu m$ for our $3\le z<4$ redshift bin. Smaller rest-$1\rm \, \mu m$ sizes in comparison to rest-UV and rest-optical sizes suggest colour gradients could be playing a role in flattening the slope. We discuss this further in Section \ref{sec:dis_z34_colgrads}.

\subsection{The Decreasing Slope Behaviour of $3<z<4$ Star-forming Galaxy Size--Mass Relation}
\label{sec:dis_z34_colgrads}
\begin{figure}
\centering
    \includegraphics[width=0.44\textwidth]{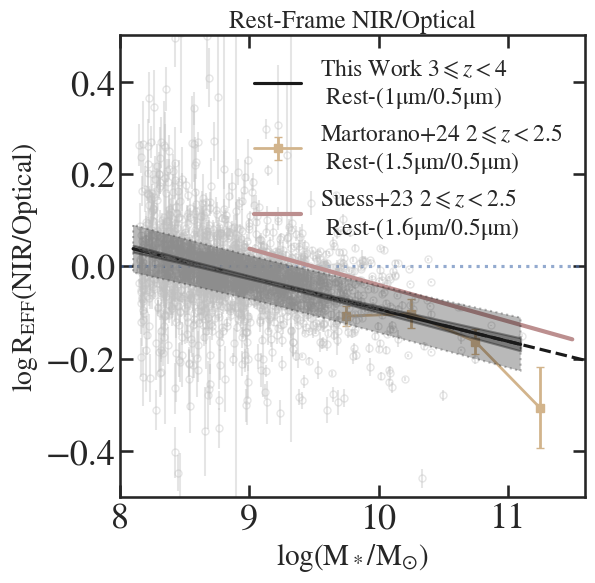}
    \caption{The ratio between the size--mass relations measured in F444W and F200W for $3\le z < 4$ (black line) corresponding to rest frame $\sim 1\rm\,\mu m$ and $\sim 0.5\rm\, \mu m$. Extrapolation of this relation to $\lms > 11$ is shown as the dashed line. Star-forming galaxies of stellar mass $\lms > 9.5 $ are $\gtrsim 10\%$ smaller in rest-$ 1\rm\,\mu m$ than in rest-$ 0.5\rm \, \mu m$ indicating the presence of colour-gradients (see Section \ref{sec:rest-nir_opitcal_comp}). Our results are in good agreement with lower redshift studies from \citet{Suess2022Rest-frameAppeared} and \citet{Martorano2024TheFields} (shown as coloured lines and squared points).}
    \label{fig:sm_ratio_f444w-f200w_z34}
\end{figure}

In this section we investigate the behaviour of the decreasing size--mass slope with increasing wavelength, for star-forming galaxies at $3\le z < 4$. 
With the large NIR coverage by JWST NIRCam, the sizes of $3\le z <4$ galaxies can be measured in wavelengths up to $\sim 1\,\rm \mu m$. NIR wavelengths are less affected by dust and thus are expected to most accurately trace the underlying mass distribution of the stellar population. 
By fitting a linear relation to the ratio of the rest-optical and rest-$\rm 1\,\mu m$ sizes (shown in Fig. \ref{fig:sm_ratio_f444w-f200w_z34}) we find that, $\lms > 10$ star-forming galaxies at $3<z<4$ are smaller by $\sim 10 \% $ in rest-$1\rm \,\mu m$ than $0.5\rm \,\mu m$. While for $\lms < 10$ star-forming galaxies, there appears to be no significant difference in size between these two wavelengths. Therefore, the more massive galaxies appear to be contributing to this flattening of the size--mass slope with increasing wavelength, as shown in Fig. \ref{fig:sm_parameters_allz_rest-lambda}. Size differences in rest-NIR and rest-optical have also been found by \citet{Suess2022Rest-frameAppeared} and \citet{Martorano2024TheFields} whose measurements are also included in Fig. \ref{fig:sm_ratio_f444w-f200w_z34}. We find a similar downwards trend to higher masses as \citet{Suess2022Rest-frameAppeared} as well as a $1\sigma$ agreement with \citet{Martorano2024TheFields} over $9.5 < \lms <10.5$.

Smaller rest-$1\rm\,\mu m$ sizes in comparison to rest-optical suggests colour gradients in $3\le z <4$ star-forming galaxies of $\lms > 10$. These colour gradients can indicate inside-out-growth, build up of bulges, or the effects of dust attenuation. Recent JWST studies have also found indications of colour gradients in $z<3$ star-forming galaxies (\citealt{Suess2022Rest-frameAppeared}, \citealt{Martorano2024TheFields}, \citealt{VanDerWel2023StellarRadii}). 
These colour-gradients could be a result of compact dust attenuation in the centre of these high-mass star-forming galaxies. Larger dust fractions in more massive galaxies have also be found by other studies (e.g. \citealt{Whitaker2017TheZ=2.5, Fudamoto2020Astronomy4.4-5.8, Magnelli2024A5, Weibel2024GalaxyObservations}) and the spatial distribution of this dust in $z>2.5$ star-forming galaxies has been found to be located centrally in some simulation (e.g. \citealt{Wu2020PhotometricSimulations}, \citealt{Marshall2021TheReionization}, \citealt{Roper2022First5}, \citealt{Popping2022The2}) and observational studies (\citealt{Tacchella2018Dust, Wang2017UVIProfiles, Nelson2016SPATIALLY1.4, Gomez-Guijarro2022AstronomyGalaxies, Fujimoto2017DemonstratingSources}).
Although, the dust distribution in high-z galaxies is still uncertain partly due limitations of current facilities (e.g. ALMA and NOEMA) that require long integration times for higher resolution data and for detecting the low surface brightness from the outskirts. Further work on obtaining large samples of high-resolution dust detection of star-forming galaxies in addition to resolved stellar population fitting is needed to understand the effects of dust on galaxy morphology.

\label{sec:rest-nir_opitcal_comp}

\subsection{Comparisons to Simulations}
\label{sec:sim_comp_dis}

In this section we discuss the findings of some simulations on the wavelength dependence on galaxy sizes and also compare the observed $5\le z < 6$ size--mass relation with predictions. 

In Section \ref{sec:results} we find that the parameters of the size--mass relation have a wavelength dependence, such as the intrinsic scatter which is larger in rest-UV wavelengths than in wavelengths at $\lambda \geqslant 0.5\rm \,\mu m$ for all our redshift bins. Similar results have been found in the FIRE simulation for $7.5 < \lms < 8$ star-forming galaxies at $z\geqslant 5$, where the scatter between half-stellar mass radii and rest B-band sizes is smaller in comparison to half-stellar mass radii and rest-UV sizes. For instance, \citet{Ma2018SimulatingSizes} find that the rest-UV sizes trace the small, bright young stellar clumps within these high-z galaxies, and these clumps do not represent the total stellar mass.

Our results also show that the slope of the $3\le z<6$ star forming size--mass relation follows a general decrease from rest-UV to rest-optical and this continues up to rest-$1\rm\,\mu m$ wavelengths, for the redshift bins that probe redder wavelengths. A similar trend has also been found in simulations for $5<z<10$ star-forming galaxies between the stellar mass range of $9<\lms<11.5$ (\citealt{Roper2022FLARESEvolution, Marshall2021TheReionization}).
These simulations find that the bulk of the stellar distribution of $\lms > 9.5$ star-forming galaxies at $z>5$ is compact and also is attenuated by centrally compact dust profiles. This dust attenuates/dims the star-forming core so that star-formation in the outskirts of these galaxies can be measured. Therefore wavelengths bluer than NIR may be biased and overestimate the galaxies "true" size (\citealt{Roper2022FLARESEvolution, Marshall2021TheReionization, Popping2022The2}). However, these simulations find negative slopes in NIR, which is in disagreement with out $3<z<4$ relations. 
Further work using bands that probe rest-NIR at $z>3$ are needed to further investigate the flattening of the size--mass relation. The JWST/MIRI instrument can provide rest-NIR observations of galaxies at $z>3$, but with reduced spatial resolution compared to the NIRCam wavelength range.

Next we compare our $5\le z <6$ size--mass relation to three simulations; Illustis-TNG (\citealt{Costantin2023ExpectationsView}), Thesan (\citealt{Shen2024TheReionization}) and Astrid (\citealt{LaChance2024THESIMULATION}), who have fit the size--mass relation for galaxies at $z=5$ and $z=6$. 
\citet{Costantin2023ExpectationsView} selects a sample of $3<z<7$ star-forming galaxies with stellar masses between $9.5< \lms <12$ from the TNG-50-1 Cosmological Simulation (the highest resolution box from the Illustris-TNG models). \citet{LaChance2024THESIMULATION} selects a similar mass range sample over the redshift range $3<z<6$ from the Astrid Simulation. Both Illustris-TNG and Astrid create mock NIRCam images that represent the CEERS observational images and measure galaxy sizes using a parametric method with the python based software,  \texttt{statmorph}. 
\citet{Shen2024TheReionization} creates noiseless mock images to measure the apparent sizes for simulated $6<z<10$ galaxies taken from the Thesan project: a suite of large-volume cosmological radiation-magneto-hydrodynamic simulations of the Epoch of Reionisation. UV and V-band 2D effective radii of galaxies are measured by taking the median distance to the centre weighted by their brightness. 
    
\begin{figure}
    \centering
    \includegraphics[width=0.44\textwidth]{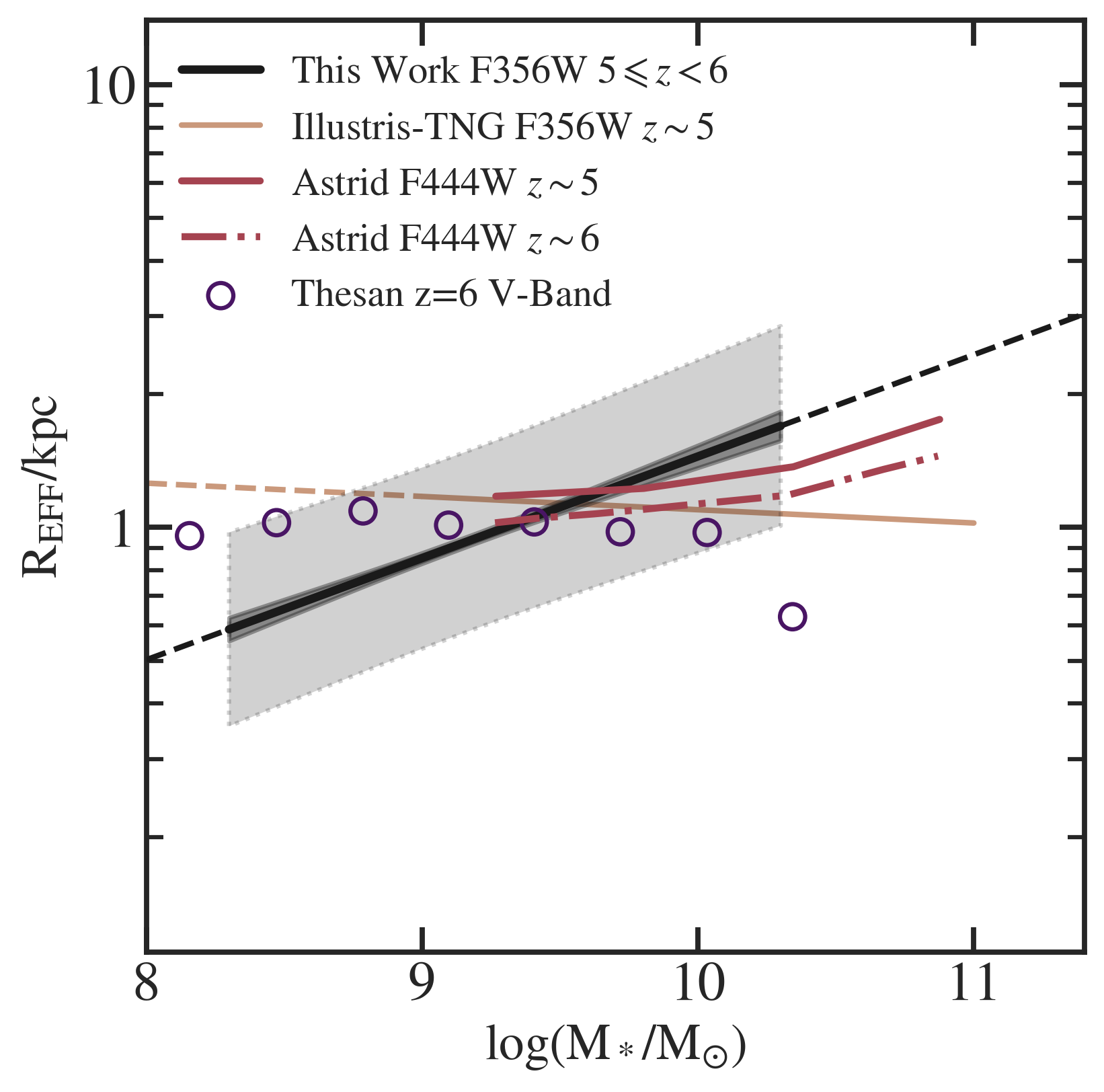}
    \caption{Our $5\le z <6$ F356W star-forming size--mass relation in comparison to three simulations: Thesan \citep{Shen2024TheReionization}, Illutris-TNG \citep{Costantin2023ExpectationsView} and Astrid \citep{LaChance2024THESIMULATION}. We find agreements between the observed and simulated $z>5$ size--mass relations over $9\le \lms < 10$. However at $\lms > 10$ there are significant disagreements, due to the compact high-mass galaxies found in simulations.}
    \label{fig:sm_z6_sim-comp}
\end{figure}

The $z>5$ size--mass relation measured in these three simulations are plotted in Fig. \ref{fig:sm_z6_sim-comp} with our $5\le z<6$ observed relation. Over the stellar mass range $9\le \lms < 10$, we are in good agreement with all simulations, given the scatter. However at $\lms > 10$, we find a steeper positive slope from observations in comparison to the simulations. Astrid finds a positive slope at these high masses but it is offset by 0.5 dex from our relation, while Illustris-TNG and Thesan find smaller sizes for high-mass galaxies, causing a down turn. 

This downturn is caused by compact massive star-forming galaxies at $z>5$. Such sources have been found in many simulations and drive the slope of size--luminosity and size--mass relations towards negative values (e.g. \citealt{Roper2022First5, Marshall2021TheReionization, Shen2024TheReionization, Popping2022The2}). The method for these compact sizes is still being studied. Although, some simulations find that the compact morphology is a result of either stars forming in dense, pristine gas at $z>10$ and stay compact until $z=4$, or diffuse $z<10$ galaxies have run away star-formation in the centres and drives them to be compact by $z=5$ (\citealt{Roper2022FLARESEvolution, Marshall2021TheReionization}). Observational evidence for the run-away star-formation mechanisms found in simulations is required. As discussed in Section \ref{sec:rest-nir_opitcal_comp} our measurements may still be affected by dust and colour gradients. The latter requires further work using IFU data, dust mapping or spatially resolved SED fitting \citep[e.g.,][]{Gimenez-Arteaga2023SpatiallyField}.
Furthermore our high-z sample is still limited to $\lms < 10.4$ star-forming galaxies. Large and deep surveys such as COMSOS-Web \citep{Cosmos-Web_Overview2024} are needed to probe these high-mass sources and constrain the high-mass end of the size--mass relation in more detail.

\section{Conclusion}
\label{sec:conclusion}

In this paper we have studied the size--mass relation for star-forming galaxies between $3\le z < 9$. By combining three fields: CEERS, PRIMER-UDS and PRIMER-COSMOS, we can fit the size--mass relation over the mass range of $8 \lesssim \lms \lesssim 10.5$. We select $z\geqslant 3$ star-forming galaxies using the SED fitting software \texttt{Eazy} and measure stellar masses as well as other physical parameters with \texttt{Bagpipes}. With the multi-wavelength model fitting tool \texttt{GalfitM}, we measure the sizes of 3529 star-forming galaxies by fitting single S\'ersic profiles over six NIRCam wide-filter bands. We use this to estimate the wavelength-dependent size--mass relation in four redshift bins: $3\le z <4$, $4\le z <5$, $5\le z <6$, and $6\le z <9$ using the Bayesian linear regression tool \texttt{Linmix}.  The main results are as followed.
\begin{enumerate}[label=(\roman*)]
    \item We find that the redshift evolution in the average sizes of $\rm M_* = 5\times 10^{10}M_{\odot}$ star-forming galaxies is well fit by a power law function of  $\log \Reff = (-0.807\pm 0.026) \times \log_{10}(1+z) +(-0.947\pm0.014)$, over $3\le z<9$. This indicates that a typical star-forming galaxy of this mass grows rapidly by a factor of 2$\times$ from $z\sim8$ to $z\sim3$ and then another factor of 3$\times$ to $z\sim0$.
    \item Through studying the evolution of the rest-optical size--mass parameters, we find that the evolution of the slope and intrinsic scatter are well fit by a constant over $3\le z<9$, where these values are $\alpha = 0.215\pm 0.009$ and $\sigma _{\log \Reff} = 0.188 \pm 0.007$. The processes that flatten the star-forming size--mass slope, compared to the relation of stellar-to-halo size, acts similarly across all redshifts.
    \item With the large NIRCam wavelength coverage in all three fields, we can parameterise the size--mass relation as a function of wavelength and redshift. We find that the slope and intrinsic scatter have a wavelength dependence, while the intercept shows no significant dependence. These parameters are largest in rest-UV in comparison to redder wavelengths. We speculate that bursty star-formation, traced by rest-UV wavelengths, causes these wavelength dependencies. 
    \item We find evidence of colour gradients in $3\le z <4$ star-forming galaxies with $\lms > 9.5$. These sources are smaller in rest-$1\rm\,\mu m$ in comparison to rest-optical (and rest-UV) wavelengths. Possible suggestions for these colour gradients are inside-out growth or bulge growth in massive galaxies at this epoch, as well as compact and central dust profiles - suggested by simulations - which attenuate the central star-forming core so that star-formation in the outskirts is measured. 
    \item Finally, we compare our $5\le z < 6$ star-forming size--mass relation to three simulations. We find a general good agreement between the observed and simulated size--mass relations over $9\le \lms <10$. However beyond this mass, the observed relation is significantly steeper than any of the simulated relations - which are flattened due to extremely compact galaxies at high masses. Such galaxies are not seen in our current sample.  
\end{enumerate} 

Future work on combining larger and deeper surveys such as COSMOS-Web are needed to constrain the size evolution of the most massive galaxies. Incorporating deeper surveys such as JADES and NGDEEP will enable robust measurements of the size--mass slope at $z>6$, determining whether or not the steepening of the size--mass relation continues towards higher redshifts. Further work in understanding the mechanism driving highly compact massive galaxies in simulations is also needed.
Finally, larger surveys with MIRI imaging are required to measure rest-NIR and rest-optical sizes at higher-z to enable the study of colour gradients and how they affect our understanding of galaxy growth. IFU or spatially resolved SED fitting can also shed light on the cause of these colour gradients. This study highlights the importance of measuring sizes in multiple wavelengths, to reduce biases and understand galaxy growth. 

\begin{acknowledgements}
The authors acknowledge the CEERS and PRIMER teams for developing their observing program with a zero-exclusive-access period.
This work is based on observations made with the NASA/ESA/CSA James Webb Space Telescope. The raw data were obtained from the Mikulski Archive for Space Telescopes at the Space Telescope Science Institute, which is operated by the Association of Universities for Research in Astronomy, Inc., under NASA contract NAS 5-03127 for JWST. 
This research is also based on observations made with the NASA/ESA Hubble Space Telescope obtained from the Space Telescope Science Institute, which is operated by the Association of Universities for Research in Astronomy, Inc., under NASA contract NAS 5–26555.
The data products presented herein were retrieved from the Dawn JWST Archive (DJA). DJA is an initiative of the Cosmic Dawn Center, which is funded by the Danish National Research Foundation under grant DNRF140.
This work has received further funding from the Swiss State Secretariat for Education, Research and Innovation (SERI) under contract number MB22.00072, as well as from the Swiss National Science Foundation (SNSF) through project grant 200020\_207349. R.G.V. acknowledges funding from the Science and Technology Facilities Council (STFC; grant code ST/W507726/1).
\newline
Software: \texttt{Astropy}, \texttt{Eazy} \citep{Brammer2008EAZY:CODE}, \texttt{Galfit}, \texttt{GalfitM} \citep{Hauler2012MegaMorph-multi-wavelengthSurveys, Vika2013MegaMorph-multiwavelengthFar}, \texttt{Grizli} \citep{BrammerG.2023Grizli}, \texttt{WebbPSF} \citep{Perrin2014UpdatedWebbPSF}, \texttt{Bagpipes} \citep{Carnall2018InferringMechanisms}

\end{acknowledgements}

\bibliographystyle{aa}
\bibliography{fixed_ref}
\newpage
\onecolumn 
\begin{appendix}

\section{Parameter Values of Size--Mass Relations}
Table \ref{tbl:sm_linear_params_allz_allbands} presents the values of the size--mass parameters measured using \texttt{Linmix}. Section \ref{sec:fitting_size-mass-relation} discusses the method used.

\begin{table*}[h]
\centering
\begin{tabular}{ccccccccc}
\hline 
\\[-5pt]

Redshift                                             & Parameter  & F115W  & F150W  & F200W  & F277W  & F356W  & F410M  & F444W  \\[3pt] \hline \hline \\[-5pt]
\multirow{5}{*}{$3\le z < 4$}                       & $\rm \alpha $ & 0.204 & 0.203 & 0.204 & 0.181 & 0.142 & 0.116 & 0.105 \\
                                                     & $\rm \sigma_{\alpha }$ & 0.012 & 0.011 & 0.010 & 0.009 & 0.009 & 0.009 & 0.009 \\
                                                     & $\rm \log_{10}A$ & 0.070 & 0.083 & 0.087 & 0.092 & 0.085 & 0.085 & 0.092 \\
($z_{\rm median} = 3.214$)                           & $\rm \sigma _{log_{10}A} $ & 0.006 & 0.005 & 0.005 & 0.004 & 0.004 & 0.004 & 0.004 \\
                                                     & $\rm \sigma_{\log \Reff}$ & 0.232 & 0.217 & 0.199 & 0.189 & 0.186 & 0.180 & 0.178 \\[3pt]
                                                     \hline \\[-5pt]

\multirow{5}{*}{$4\le z <5$}                        & $\rm \alpha $ & 0.213 & 0.220 & 0.218 & 0.231 & 0.200 & 0.159 & 0.151 \\
                                                     & $\rm \sigma_{\alpha }$ & 0.015 & 0.014 & 0.013 & 0.012 & 0.012 & 0.012 & 0.012 \\
                                                     & $\rm \log_{10}A$ & -0.035 & -0.031 & -0.012 & -0.012 & -0.002 & 0.005 & 0.014 \\
($z_{\rm median} = 4.428$)                           & $\rm \sigma _{\log_{10}A} $ & 0.007 & 0.007 & 0.006 & 0.005 & 0.005 & 0.006 & 0.006 \\
                                                     & $\rm \sigma_{\log \Reff}$ & 0.225 & 0.211 & 0.191 & 0.174 & 0.171 & 0.171 & 0.168  \\[3pt]
                                                     \hline \\[-5pt]
{\multirow{5}{*}{$5\le z < 6$}}                     & $\rm \alpha $ & 0.303 & 0.294 & 0.288 & 0.259 & 0.231 & 0.170 & 0.173 \\
                                                     & $\rm \sigma_{\alpha}$ & 0.033 & 0.031 & 0.030 & 0.028 & 0.028 & 0.029 & 0.029 \\
                                                     & $\rm \log_{10}A$ & -0.115 & -0.109 & -0.105 & -0.082 & -0.070 & -0.023 & -0.025 \\
($z_{\rm median} = 5.278$)                           & $\rm \sigma _{\log_{10}A} $ & 0.015 & 0.014 & 0.013 & 0.012 & 0.013 & 0.013 & 0.013 \\
                                                     & $\rm \sigma_{\log \Reff}$ & 0.227 & 0.214 & 0.210 & 0.197 & 0.193 & 0.188 & 0.191 \\[3pt]
                                                     \hline \\[-5pt]
{\multirow{5}{*}{$6\le z < 9$}}                     & $\rm \alpha $ & 0.215 & 0.215 & 0.215 & 0.215 & 0.215 & 0.215 & 0.215 \\
                                                     & $\rm \sigma_{\alpha }$ & 0.001 & 0.001 & 0.001 & 0.001 & 0.001 & 0.001 & 0.001 \\
                                                     & $\rm \log_{10}A$ & -0.176 & -0.184 & -0.170 & -0.153 & -0.149 & -0.133 & -0.120 \\
($z_{\rm median} = 6.419$)                           & $\rm \sigma _{log_{10}A} $ & 0.028 & 0.027 & 0.024 & 0.019 & 0.017 & 0.019 & 0.018 \\
                                                     & $\rm \sigma_{\log \Reff}$ & 0.259 & 0.247 & 0.222 & 0.178 & 0.158 & 0.165 & 0.163 \\[3pt]
                                                       \hline
\end{tabular}

\caption{Best-fit parameters from \texttt{Linmix} fits to the size mass equation $\log \rm \Reff = \alpha \log(M_*/M_0) + logA$ in four redshift bins: $3\leqslant z < 4$, $4\leqslant z < 5$, $5\leqslant z < 6$, $6\leqslant z < 9$. Here $\alpha$ is the slope, $\rm M_0$ ($\rm \log M_0 = 9$) is the intercept and $\rm \sigma_{\log \Reff}$ is the intrinsic scatter (see section Section~\ref{sec:fitting_size-mass-relation} for details).}
\label{tbl:sm_linear_params_allz_allbands}
\end{table*}

\section{Derivation of \texttt{GalfitM} S\'ersic Model Uncertainties}
\begin{figure*}
    \centering
    \includegraphics[width=1\textwidth]{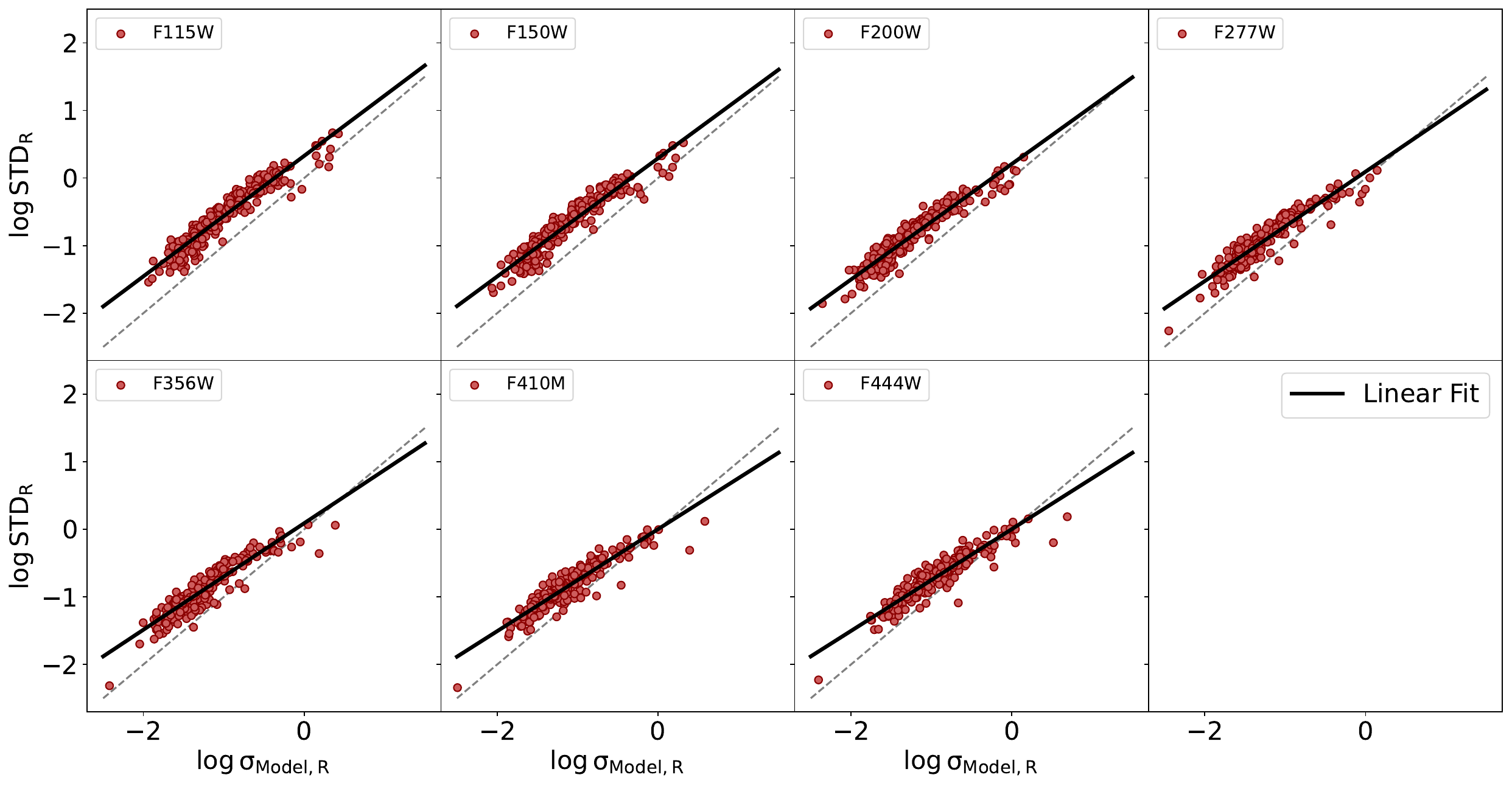}
    \caption{Standard deviation in the effective radius from \texttt{GalfitM} measurements based on simulated images (see Section \ref {sec:galfit-err-estimates}). The log of the standard deviation of $\Reff$ ($\log \rm STD_R$) is related to the log of the error on the effective radius of the input model via a linear relationship: $\log \rm STD_R = a*\log \sigma_{\Reff} + b$. The black solid line shows the fitted relationship in each band and the equation for each fit is included in Table \ref{tbl:galfitm-sim-err-run_linear-params}. Dashed lines show the one-to-one relation.}
    \label{fig:simulated-galfitm-errors_std-err}
\end{figure*}

As discussed in Section \ref{sec:galfit-err-estimates}, we found that the uncertainties on the S\'ersic parameters calculated by \texttt{GalfitM} are underestimated. Thus, we run a Monte-Carlo simulation on simulated images to calculate a correction factor. Using ds9, we selected 150 blank patches in the CEERS \texttt{grizli} reduced images. Then 250 random models selected from our \texttt{GalfitM} run on CEERS images, were separately inputted into each of the 150 blank patches. \texttt{GalfitM} is run on each of these models in the same way as described in Section \ref{sec:galfitm}. With the 150 outputs of each model, we calculated a standard deviation in the effective radius. We found no correlation between this standard deviation and the signal-to-noise of the models. However, we did find a correlation between the log of the  standard deviation from the Monte-Carlo simulation ($\rm \log STD_{R}$) and the log of the effective radius uncertainty returned by \texttt{GalfitM} ($\log \sigma _{\Reff}$, from the input model). Between these two parameters, we fit a linear relation and this is presented in Fig. \ref{fig:simulated-galfitm-errors_std-err} and Table \ref{tbl:galfitm-sim-err-run_linear-params}. These linear relations are applied to correct the uncertainties from \texttt{GalfitM}.

\begin{table*}[h]
\centering
\begin{tabular}{ccc}
\hline \\[-4pt]
Band & Slope, a & Intercept, b \\
\\[-4pt]
\hline
\hline  \\[-4pt]
f115w & 0.8917 & 0.3282 \\
f150w & 0.8748 & 0.2918 \\
f200w & 0.8551 & 0.2079 \\
f277w & 0.8100 & 0.0946 \\
f356w & 0.7884 & 0.0909 \\
f410m & 0.7547 & 0.0025 \\
f444w & 0.7535 & 0.0034 \\[3pt]
\hline
\end{tabular}
\caption{The linear parameters from $\rm \log STD_R = a*\log \sigma_{
\Reff} + b$ where $\rm STD_R$ is the standard deviation calculated from our Monte-Carlo simulation and $\sigma_{\Reff}$ is the uncertainty on the effective radius from the input model, as returned by \texttt{GalfitM}
 (See Section \ref{sec:galfit-err-estimates} for details).}
\label{tbl:galfitm-sim-err-run_linear-params}
\end{table*}

\end{appendix}
\end{document}